\pdfoutput=1
\documentclass[iop]{emulateapj}
\usepackage{natbib}
\usepackage{amsmath}
\usepackage{verbatim}
\usepackage{float}

\usepackage{subfigure}

\begin{document}

\title{\textbf{Near-Earth Asteroid Satellite Spins Under Spin-Orbit Coupling}}
\date{}
\author{Shantanu P. Naidu\altaffilmark{1}, Jean-Luc Margot\altaffilmark{1,2}}

\altaffiltext{1}{Department of Earth, Planetary, and Space Sciences, University of California, Los Angeles, CA 90095, USA}
\altaffiltext{2}{Department of Physics and Astronomy, University of California, Los Angeles, CA 90095, USA}

\begin{abstract}
We develop a fourth-order numerical integrator to simulate the coupled
spin and orbital motions of two rigid bodies having arbitrary mass
distributions under the influence of their mutual gravitational
potential. We simulate the dynamics of components in
well-characterized binary and triple near-Earth asteroid systems and
use surface of section plots to map the possible spin configurations
of the satellites.  For asynchronous satellites, the analysis reveals
large regions of phase space where the spin state of the satellite is
chaotic.  For synchronous satellites, we show that libration
amplitudes can reach detectable values even for moderately elongated
shapes.  The presence of chaotic regions in the phase space has
important consequences for the evolution of binary asteroids.  It may
substantially increase spin synchronization timescales, explain the
observed fraction of asynchronous binaries, delay BYORP-type
evolution, and extend the lifetime of binaries. The variations in spin
rate due to large librations also affect the analysis and
interpretation of lightcurve and radar observations.
\end{abstract}

\section{Introduction}
Binary near-Earth asteroids (NEAs) are numerous in the asteroid
population.  Both radar and lightcurve data have shown that
$\sim16\%$ of NEAs larger than $\sim$200~m diameter have
satellites~\citep{pravec99,margot02,pravec06}.  It is now widely
accepted that binary NEAs form by a spin-up process \citep{margot02}
and that the specific spin-up mechanism is the YORP torque
\citep{rubincam00}.  Binary NEA systems exhibit interesting
post-fission and spin-orbit
dynamics~\citep[e.g.][]{ostro06,scheeres06,fahnestock08,mcmahon2013}
that profoundly affect their evolution~\citep[e.g.][]{jacobson11,
fang12, jacobson14}, but the range of dynamical regimes has not been
fully explored.

In this paper, we develop a method for simulating the coupled spin and
orbital motions of two rigid bodies with arbitrary mass
distributions. This technique is significantly faster than a similar
implementation by \citet{fahnestock06}, because in our implementation
the computationally expensive volume integrals over the two bodies are
computed only once before the integration, as opposed to once per time
step.  We use our technique to perform a survey of the dynamics of all
well-characterized binary NEA systems and map the range of dynamical
behaviors, including the spin configurations of asteroid satellites.
These results provide important insights for modeling the physical
properties of binaries and for understanding the long term evolution
of the binary systems.

The sample of well-characterized binaries includes all NEA systems
with known estimates of system mass, semi-major axis, eccentricity,
and component sizes.  In practice, only systems observed with radar
fall in this class.  Over 35 binary NEAs have been observed with
radar, but only about ten have sufficient data to yield mutual orbits
and component size estimates.  We apply our technique to these
systems.

Sections~\ref{sec:num_int} and \ref{sec:econ} describe the
implementation of our coupled spin-orbit integrator and cover energy
and angular momentum conservation properties.
Section~\ref{sec:notation} explains different kinds of satellite spin
librations and sets up the notation used in subsequent sections. In
section~\ref{sec:lib}, we examine the spin-orbit coupling effect and
compare numerical and analytical estimates of libration
amplitudes. Section~\ref{sec:sos} introduces surface of section plots
which are used to identify resonant, chaotic, and non-resonant
quasi-periodic trajectories. We examine the trajectories and spin
configurations of satellites in well-characterized binary and triple
near-Earth asteroid systems in section~\ref{sec:systems} and show that
large chaotic zones exist in the phase space of known asynchronous
satellites. We also compute libration amplitudes for synchronous
satellites.  We discuss implications of the results in
section~\ref{sec:implications}.

\section{Numerical Integration}
\label{sec:num_int}
We numerically investigate the coupled spin and orbital dynamics of
two extended rigid objects under their mutual gravitational
influence. We neglect the translational motion of the system
barycenter and use the 6 first-order differential equations of motion
(EOMs) derived by \citet{maciejewski95}.  Here we express these EOMs
in the body-fixed frame of the primary:

\begin{align}
&\dot{\boldsymbol{P}}=\boldsymbol{P}\boldsymbol{\times}\boldsymbol{\Omega}_1-\frac{\partial V}{\partial \boldsymbol{R}},    &    & \dot{\boldsymbol{R}}=\boldsymbol{R}\boldsymbol{\times}\boldsymbol{\Omega}_1+\frac{\boldsymbol{P}}{m}\notag,\\
&\dot{\boldsymbol{\Gamma}}_2=\boldsymbol{\Gamma}_2\boldsymbol{\times}\boldsymbol{\Omega}_1+\boldsymbol{\mu}_2,    &    & \dot{\boldsymbol{\Gamma}}_1=\boldsymbol{\Gamma}_1\boldsymbol{\times}\boldsymbol{\Omega}_1+\boldsymbol{\mu}_1\notag,\\
&\dot{S}=S\hat{\Omega}_2-\hat{\Omega}_1S,    &    & \dot{S}_1=S_1\hat{\Omega}_1.
\label{eq:eom}
\end{align}

\noindent Here $\boldsymbol{R}$ and $\boldsymbol{P}$ are the relative
position and linear momentum vectors of the secondary with respect to
the primary, respectively, $m=m_pm_s/(m_p+m_s)$ is the reduced mass of
the system, where $m_p$ and $m_s$ are the masses of the primary and
secondary, respectively, $V$ is the mutual gravitational potential,
$\boldsymbol{\mu}$'s are the torque vectors acting on the two
components, $\boldsymbol{\Omega}$'s are their angular velocity
vectors, and $\boldsymbol{\Gamma}$'s are their angular momentum
vectors.  Subscripts 1 and 2 denote quantities that refer to the
primary and the secondary, respectively. Further, $S$ and $S_1$ are
attitude rotation matrices: the former mapping from the secondary
frame to the primary frame, and the latter mapping from the primary
frame to the inertial frame. A hat ( $\hat{}$ ) symbol above a vector
specifies an operator that maps a 3-vector (e.g.,
$\boldsymbol{v}=[v_x,v_y,v_z]$) to an antisymmetric 3x3 matrix, as
follows:
\begin{equation}
\hat{v}=
\begin{bmatrix}
0    & -v_z & v_y  \\
v_z  & 0    & -v_x \\
-v_y & v_x  & 0    
\end{bmatrix}.
\end{equation}
The term $\partial{V}/\partial{\boldsymbol{R}}$ is the gradient of the
mutual gravitational potential, which is the gravitational force
(vector) between the two components. All vectors in
equations (\ref{eq:eom})
are expressed in the body-fixed frame of the primary.  However, when
computing $\hat{\Omega}_2$, one must express $\boldsymbol{\Omega}_2$ in
the body-fixed frame of the secondary.

The gravitational force and torques are computed at each time step
according to the method detailed in \citet{ashenberg07}. They are
functions of $\boldsymbol{R}$, $S$, and the inertia integrals of the
two bodies.  The inertia integrals encode the mass distribution
information of a body and are of the form:
\begin{equation}
I_{x^py^qz^r}=\int_B x^py^qz^rdm,
\end{equation}
\noindent where $dm$ is a mass element of body $B$ at body-fixed
coordinates ($x$, $y$, $z$) and the integral is a volume integral over
the entire body. The body-fixed coordinate system is aligned with the
principal axes and its origin is at the center of mass of $B$. The
exponents $p$, $q$, and $r$ are either 0 or positive integers, such
that $p+q+r>0$. We use inertia integrals up to fourth order in the
integrations, where the order of an inertia integral is given by the
sum of exponents, i.e., $p+q+r$.  The inertia integrals depend only on
the mass distribution of the object and remain constant throughout the
integration, so we compute them only once before the integration. At
each time step, current values of $\boldsymbol{R}$ and $S$ from the
integrator are passed as arguments to the modules that compute force
and torques.

Because detailed 3D shape models of both the primary and secondary are
generally not available and their density distributions are unknown,
we model the primary and secondary as triaxial ellipsoids 
(semi-axes $a$, $b$, and $c$) with uniform density in this paper.
These restrictions can be easily lifted as knowledge progresses.  The
uniform-density ellipsoid assumption simplifies the computation of
inertia integrals~\citep{boue09}: they are zero for odd $p$, $q$, or
$r$, and the non-zero integrals are simple functions of principal
moments of inertia. The fourth-order inertia integrals can be found in
\citet{boue09}, and the non-zero second-order inertia integrals are
listed below:
\begin{align}
&I_{x^2}=\int x^2dm=\frac{(-A+B+C)}{2},\notag\\
&I_{y^2}=\int y^2dm=\frac{(A-B+C)}{2},\notag\\
&I_{z^2}=\int z^2dm=\frac{(A+B-C)}{2}.
\end{align}
\noindent Here, 
$A\leq B\leq C$ are the principal moments of inertia of the object
about the $x$, $y$, and $z$ axes, respectively.

We use the Cash-Karp method \citep{cash90} to integrate
equations~(\ref{eq:eom}). It is a fifth-order Runge-Kutta integrator
with adaptive stepsize control which uses an embedded fourth-order
Runge-Kutta formula to compute errors. We use the implementation
provided by \citet{nr2} and set the fractional error tolerance to
$10^{-15}$.

In all simulations, we assume a planar system, i.e., both bodies are
in principal axis rotation about their $z$ (shortest) axes and their
equatorial planes are aligned with the mutual orbit at all times. We
start all simulations at the pericenter of the osculating mutual orbit
and with the longest axis of each body pointing towards each other.
The system parameters and initial osculating mutual orbital parameters
for all simulations are given in Table~\ref{tab:parameters}.

\section{Energy and Angular Momentum Conservation}
\label{sec:econ}
\begin{figure}
  \subfigure{\plotone{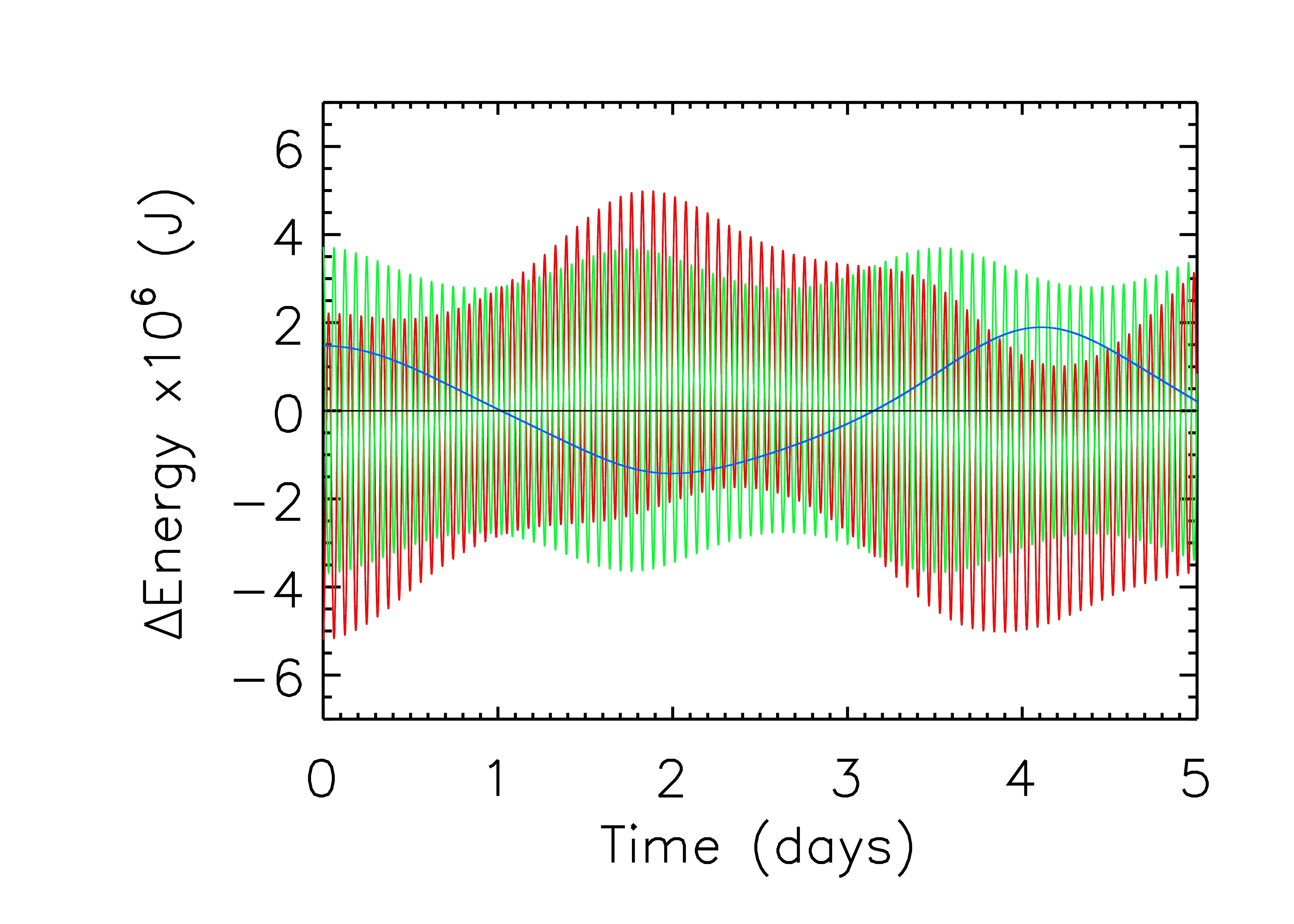}}
  \subfigure{\plotone{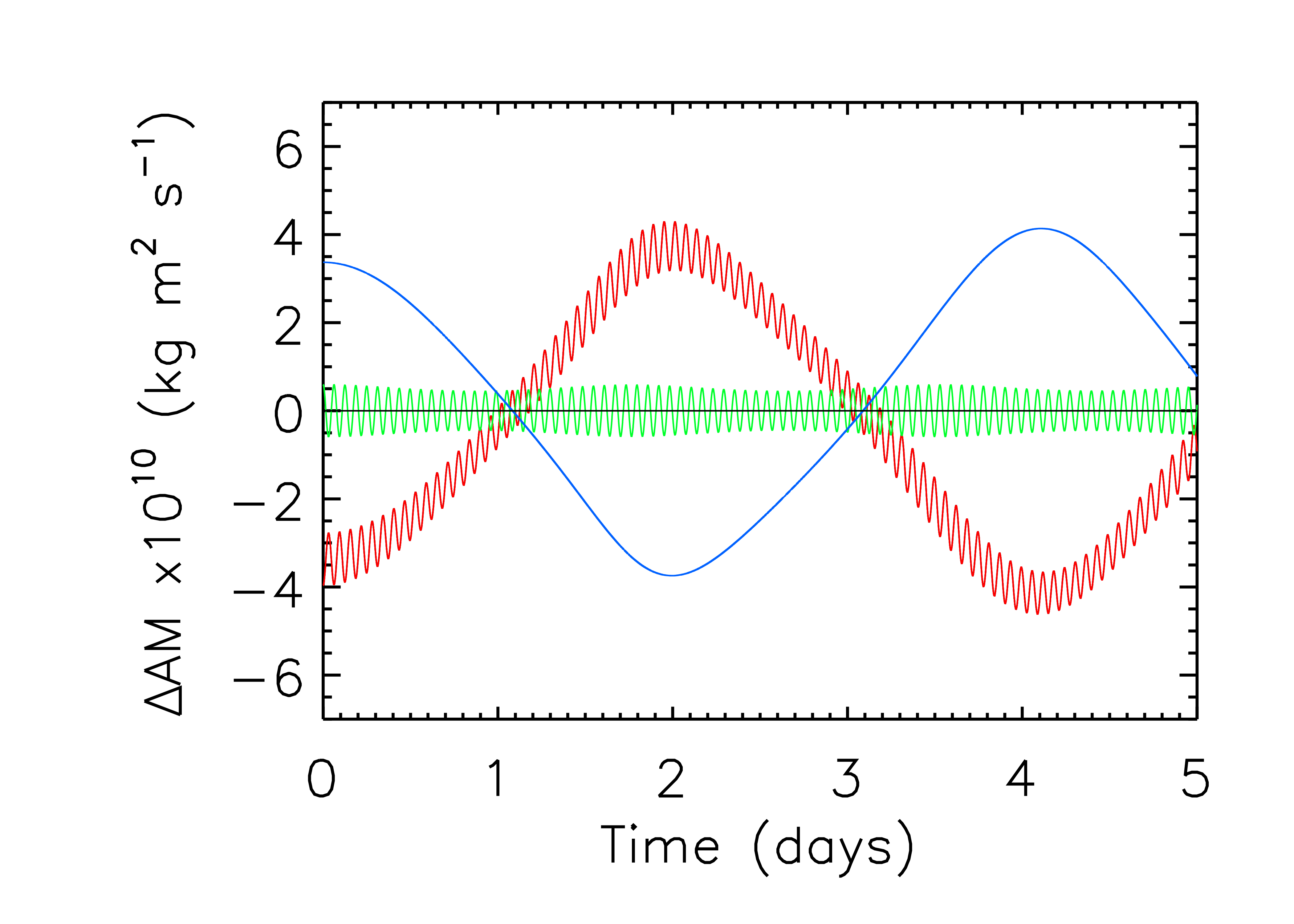}}

\caption{Energy (top) and angular momentum (bottom) variations over
  the course of five days for a typical binary NEA (first entry in
  Table 1).  Lines of different colors represent quantities associated
  with the mutual orbit (red), the primary spin (green), the secondary
  spin (blue), and the sum of all three (black).  }
\label{fig:econ}
\end{figure}

In this section, we describe results of tests designed to evaluate the
energy and angular momentum conservation properties of the integrator.
Figure~\ref{fig:econ} illustrates a representative test run with the
parameters given in the first line of Table~\ref{tab:parameters}.  For
this test case, we used triaxial ellipsoids with principal axis
half-lengths of $a=600$, $b=500$, and $c=400$ m for the primary, and
$a=252$, $b=229$, $c=190$ m for the secondary.  The initial spin periods
of the primary and the secondary are 2.775 h and 32.59 h,
respectively.

The total energy is conserved at a level of $10^{-2}$ Joules per year,
which is about $10^{-11}$ times the mean orbital energy and less than
$10^{-8}$ times the magnitude of the energy exchanged between the
binary components and the mutual orbit.

Angular momentum is conserved at a level of 220~kg~m$^2$~s$^{-1}$ per
year, which is less than $10^{-11}$ times the total angular momentum
of the system ($\sim~10^{14}$~kg~m$^2$~s$^{-1}$) and less than
$10^{-8}$ times the angular momentum exchanged between the component
spins and the mutual orbit.

\begin{deluxetable*}{llrrrrcccrr}
\tablecaption{Simulation Parameters}

\tablehead{ \multicolumn{2}{c}{} & \multicolumn{2}{c}{Primary} &
\multicolumn{5}{c}{Secondary} & \multicolumn{2}{c}{Mutual Orbit} \\
\colhead{Fig.} & \colhead{Object} & \colhead{$R_p$} & \colhead{$\rho_p$} & \colhead{$c$} & \colhead{$ab$} &
\colhead{$a/b$} & \colhead{$\omega_0$} & \colhead{$\rho_s$} & \colhead{a} &
\colhead{e}\\ 
\colhead{} & \colhead{} & \colhead{m} & \colhead{kg m$^{-3}$} &
\colhead{m} & \colhead{m$^2$} & \colhead{} & \colhead{kg m$^{-3}$} &
\colhead{m} & \colhead{}}

\startdata 

\ref{fig:econ}       &  Test             & 493  & 1581 & 190 & 57600 & 1.10    & 0.53    & 2618 & 3300 & 0.05\\ 
\ref{fig:libvsq}     &  (1991 VH)$^d$    & 600  & 1581 & 190 & 57600 & 1.50    & 1.07    & various & 3300 & 0.05\\
\ref{fig:ss_reg}     &  (1991 VH)$^l$    & 600  & 1581 & 190 & 57600 & 1.01    & 0.17    & 2618 & 3300 & 0.05\\ 
\ref{fig:ss_overlap} &  (1991 VH)$^m$    & 600  & 1581 & 190 & 57600 & 1.06    & 0.42    & 2618 & 3300 & 0.05\\ 
\ref{fig:vh_ss}      &  (1991 VH)        & 600  & 1581 & 190 & 57600 & 1.50    & 1.07    & 2618 & 3300 & 0.05\\ 
\ref{fig:yt1_ss}     &  (2003 YT1)       & 550  & 2712 &  88 & 11025 & 1.30    & 0.88    & 3248 & 3930 & 0.18\\ 
\ref{fig:dc_ss}      &  (2004 DC)        & 180  & 1461 &  26 & 900   & 1.30    & 0.88    & 2000 &  750 & 0.30\\ 
\ref{fig:all_lib}    &  (Didymos)        & 400  & 1955 &  65 & 5625  & various & various & 2252 & 1180 & 0.04\\ 
\ref{fig:all_lib}    &  (2000 DP107)     & 400  & 1791 & 100 & 22500 & various & various & 2122 & 2692 & 0.03\\ 
\ref{fig:all_lib}    &  (2002 CE26)      & 1750 &  966 & 100 & 22500 & various & various & 1454 & 4870 & 0.025\\ 
\ref{fig:all_lib}    &  (2001 SN263\#1)  & 1300 &  996 & 190 & 52900 & various & various & 2320 & 3800 & 0.016\\ 
\ref{fig:all_lib}    &  (1994 CC\#1)     & 310  & 2076 & 48  & 3249  & various & various & 8870 & 1730 & 0.002\\
\ref{fig:all_lib}    &  (1999 KW4)       & 659  & 1970 & 190 & 51076 & 1.30    & 0.88    & 3321 & 2548 & 0.0004\\
\ref{fig:all_lib}    &  (1996 FG3)       & 850  & 1300 & 200 & 60025 & 1.30    & 0.88    & 1592 & 2535 & 0.07\\ 

\enddata

\tablecomments{
Most physical and orbital characteristics of binary and triple NEAs
are adopted or derived from~\citet{fang12}.  Parameters for 1996~FG3
are from \citet{scheirich15}. Mass and radius uncertainties are
$\sim$10\% and $\sim$20\%, respectively.  See text for prescription
for $a$, $b$, and $c$ values.  The first column reports the number of
the figure illustrating the corresponding results.  ``Object''
indicates the asteroid name or designation.  The next two columns list
parameters related to the primary: $R_p$ and $\rho_p$ are the
equivalent radius and mass density of the primary. The next five
columns describe parameters related to the secondary (assumed to be an
ellipsoid with semi-axes $a$, $b$, and $c$).  With our choice of
simulation parameters (Section~\ref{sec:systems}), it is convenient to
tabulate the quantities $c$, $ab$, and the elongation $a/b$.  The
fourth parameter describing the secondary, $\omega_0$, is related to
the secondary elongation (Refer to section~\ref{sec:notation} for
definition). The fifth parameter is the mass density $\rho_s$.  The
last two columns give the initial osculating semimajor axis and
eccentricity of the mutual orbit.  For testing purposes, we use
several modified versions of binary NEA 1991 VH: $^d$ for various
densities, $^l$ for low secondary elongation, $^m$ for moderate
secondary elongation.}

\label{tab:parameters}
\end{deluxetable*}

\section{Notation and Libration Concepts}
\label{sec:notation}

\begin{figure}
\plotone{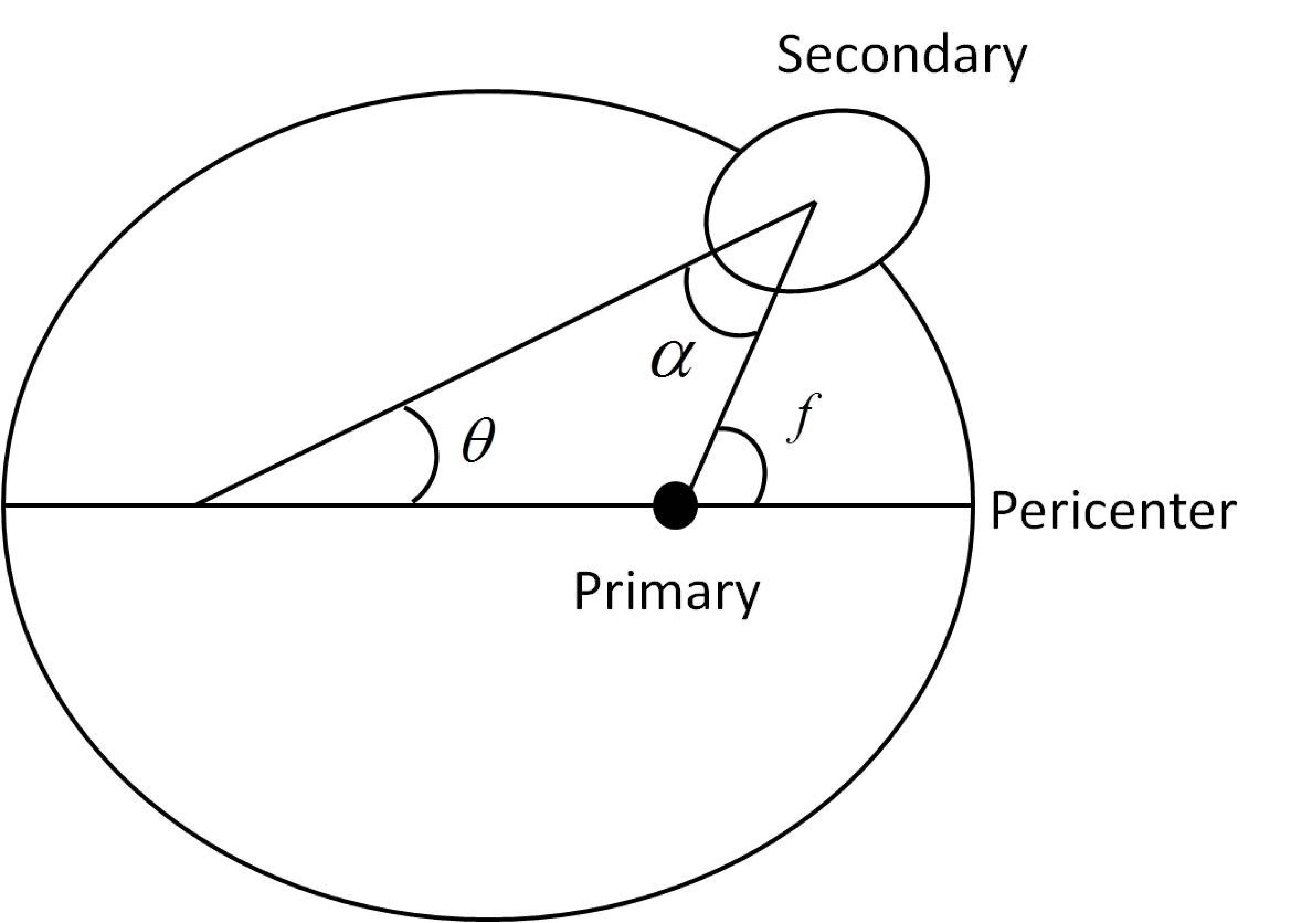}
\caption{Illustration of the osculating mutual orbit and notation for angles.}
\label{fig:angles}
\end{figure}

Figure~\ref{fig:angles} illustrates the various angles used throughout
the paper.  $\theta$ is the angle between the secondary's long axis
and the line of apsides of the the osculating mutual orbit, and
$\dot{\theta}$ is its time rate of change.  If the apsidal precession
rate were zero, $\dot{\theta}$ would correspond to the spin rate of
the satellite.  The instantaneous values at pericenter are denoted
with a subscript $p$: $\theta_p$, $\dot{\theta_p}$.  The angle
$\theta$ is related to the angle between the satellite's long axis and
the primary-secondary line, $\alpha$, by $\theta+\alpha=f$, where $f$
is the true anomaly of the mutual orbit.  At pericenter, $f=0$, so
$\theta_p=-\alpha$.

Oscillations of the secondary orientation with respect to the
primary-to-secondary line are called {\em librations}.  In order to
illustrate librations, let us first examine a situation in which the
amount of angular momentum exchanged between the spin of the secondary
and the mutual orbit is negligible.  In this situation, we can treat
the spin and orbit to be decoupled.  A common approach to analyze the
spin of the secondary is to assume the secondary to be a triaxial
ellipsoid on a fixed Kelperian mutual orbit about a spherical primary.
The secondary spin is affected by the gravitational torques exerted by
the primary\citep[e.g.,][]{murray99}.  In this situation, the angle
$\alpha$ is the sum of {\em free, forced, and optical} libration
angles.  {\em Free libration} is easiest to understand in the case of
a circular mutual orbit and a synchronously spinning secondary, i.e.,
a secondary whose average spin rate is equal to the mutual orbit mean
motion.  The minimum energy configuration for this system is for the
long axis of the secondary to always point towards the primary, such
that its instantaneous spin rate is always equal to the mean
motion. If the secondary is disturbed from this configuration, its
long axis oscillates about the primary-secondary line due to torques
exerted by the primary on the elongated secondary.  This oscillation
is called free libration and its frequency depends on the shape of the
secondary and the mutual orbit parameters.  Generally, free libration
damps out on short timescales due to tidal friction \citep{murray99}.

If the mutual orbit is eccentric, the secondary exhibits optical and
forced librations about the primary-secondary line even if the free
libration is damped out. {\em Optical libration} is the torque-free
oscillation of the long axis of a uniformly spinning secondary about
the primary-secondary line.  This oscillation would occur even in the
case of a spherical secondary as the orbital velocity varies over the
course of the orbit.  We use $\phi$ to represent the component of
$\alpha$ that is due to optical librations.  The amplitude of optical
libration depends only on the shape of the mutual orbit and is
$\sim2e$, where $e$ is the eccentricity of the mutual orbit
\citep{murray99}.  In the case of an elongated secondary, the primary
exerts a periodically reversing torque on it due to the misalignment
of the secondary long axis from the primary-secondary line, which
results in an oscillation of the secondary about uniform rotation
called {\em forced libration}.  We use $\gamma$ to represent the
component of $\alpha$ that is due to forced librations.  Forced and
optical librations have the same frequency (equal to the mean motion).
They are in phase if $\omega_0=\sqrt{3(B-A)/C}<1$ and 180$^\circ$ out
of phase if $\omega_0 > 1$.  We use $\psi$ to represent the sum of
forced and optical librations, i.e., $\psi=\gamma+\phi$, and $\psi_A$
to represent the libration amplitude.

For most binary near-Earth asteroid systems, a decoupled framework
does not accurately capture the system dynamics.  Nevertheless, even
in the fully coupled problem around an axially symmetric primary, the
secondary exhibits libration behavior similar to the free, forced, and
optical librations of the decoupled spin problem. There are two modes
of libration in the coupled spin-orbit problem, which we call the {\em
relaxed mode} and {\em excited mode} of libration. The relaxed mode
has the same frequency as the orbital frequency, similar to
forced+optical libration in the decoupled spin problem. The excited
mode of libration has a different frequency that depends on the shape
of the secondary. This libration mode is similar to free libration in
the decoupled spin problem.  By exploring a range of initial
conditions, we can minimize the excited-mode librations so that its
amplitude is close $0^\circ$, leaving the secondary librating in the
relaxed mode. The relaxed mode disappears only when the system is in
an equilibrium state, i.e., when the mutual orbit is circular and the
long axis of the secondary always points towards the primary.

For systems in which the exchange of angular momentum in the system is
small, the coupled spin-orbit problem approaches the decoupled problem
and the relaxed-mode and excited-mode librations become similar to the
forced+optical and free librations, respectively.  Because most of our
simulations include some amount of spin-orbit coupling, we use the
relaxed/excited mode terminology as opposed to the free/forced mode
terminology of the decoupled problem.

\section{Effect of Spin-Orbit Coupling on Libration}
\label{sec:lib}

In this section, we study the effects of spin-orbit coupling on the
relaxed-mode libration amplitude of the secondary.  Under the
assumptions of a fixed orbit around a spherical or point-mass primary,
the amplitudes of forced+optical librations ($\psi_A$) in the
decoupled case can be estimated with ~\citep[e.g.,][]{tiscareno09}:
\begin{equation}
\psi_A=\frac{2e}{\omega_0^2-1},
\label{eq:lib}
\end{equation}
\noindent where $e$ is the eccentricity of the mutual orbit and
$\omega_0=\sqrt{3(B-A)/C}$ 
is the natural frequency of libration of the satellite normalized by
the mean motion ($n$) of the mutual orbit.
In the coupled problem, the amplitude of the librations depends on the
primary-to-secondary mass ratio, which we quantify with our fully
coupled spin and orbit integrator.

\begin{figure}
\plotone{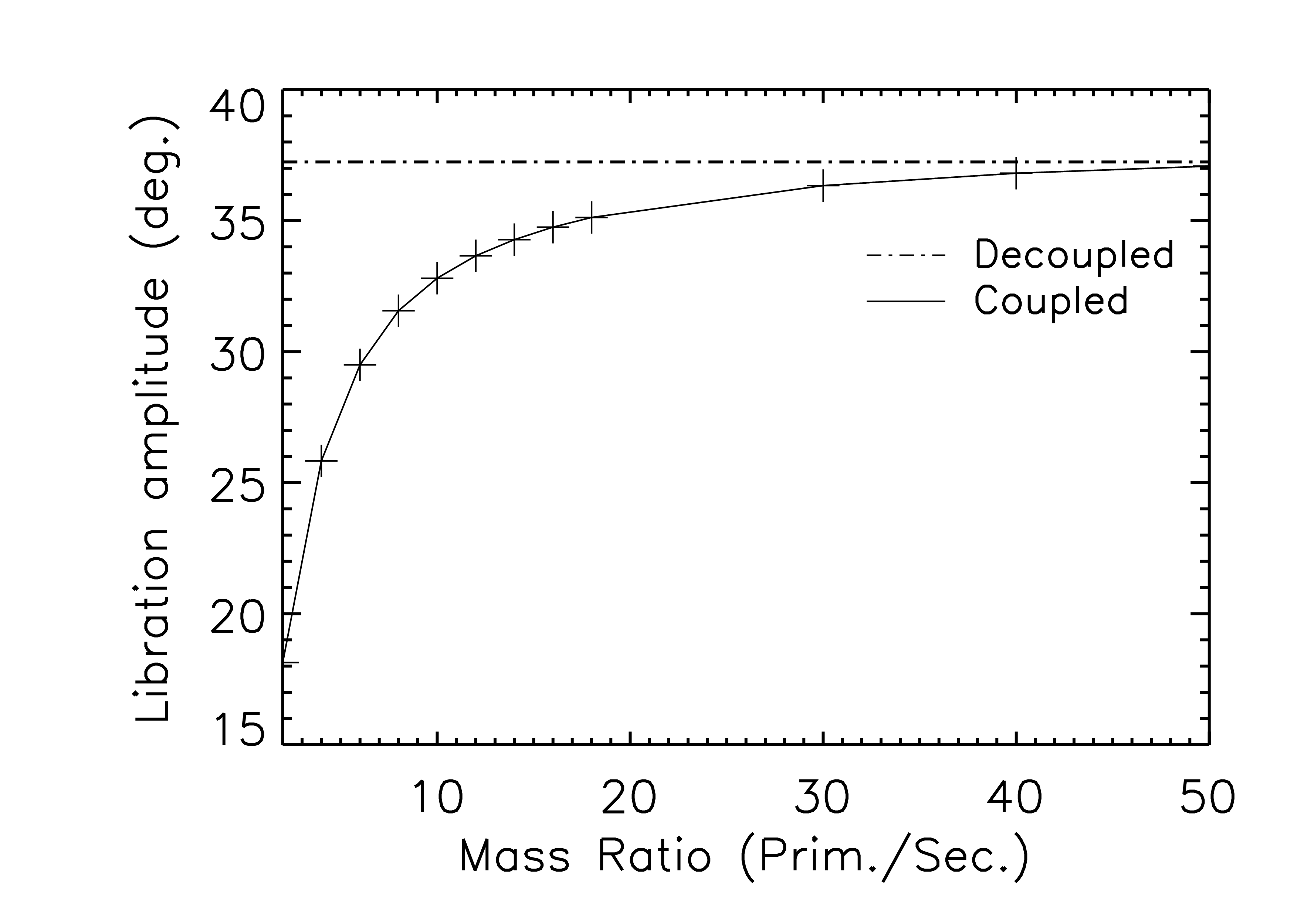}
\caption{Influence of spin-orbit coupling on relaxed-mode libration
  amplitude.  Plus symbols (connected by a solid line) show amplitudes
  of relaxed-mode libration as a function of primary-to-secondary mass
  ratio.  Dot-dashed line shows corresponding analytical estimates of
  forced+optical libration computed using equation~(\ref{eq:lib}).
  System parameters for this simulation are based on asteroid 1991 VH
  and are given in the second row of
  Table~\ref{tab:parameters}.  
  The elongation of the secondary, $a/b=1.5$, corresponds to $\omega_0=1.07$.
}
\label{fig:libvsq}
\end{figure}

Figure~\ref{fig:libvsq} shows results of simulations in which we vary
the primary-secondary mass ratio for a binary system based on NEA 1991
VH (row 2 of Table~\ref{tab:parameters}, nominal mass ratio $\approx
12$).  We vary the density of the secondary while keeping other shape
parameters constant, and we use initial conditions that make the
excited-mode libration amplitude $\sim0^\circ$.  The corresponding
analytical estimates (equation (\ref{eq:lib})) yield
$\psi_A=37.2^\circ$ for $\omega_0=1.07$ and $e=0.05$.  At low values
of the primary-to-secondary mass ratio, the libration amplitudes are
considerably smaller than the analytical estimate, suggesting that
spin-orbit coupling tends to damp libration amplitudes.

\section{Surface of Section}
\label{sec:sos}

It is useful to visualize the dynamics with surface of section plots
similar to those in \citet{wisdom84}.  At every pericenter passage of
the secondary, we plot the angle between the long axis and the line of
apsides of the mutual orbit, $\theta_p$, against its time derivative,
$\dot{\theta_p}$, normalized by the mean motion, $n$.
In order to identify pericenter passage, we
use Keplerian elements to describe the osculating
mutual orbit at each time step. These elements vary on timescales
shorter than the orbital period because the orbit is not Keplerian.

\begin{figure}
\plotone{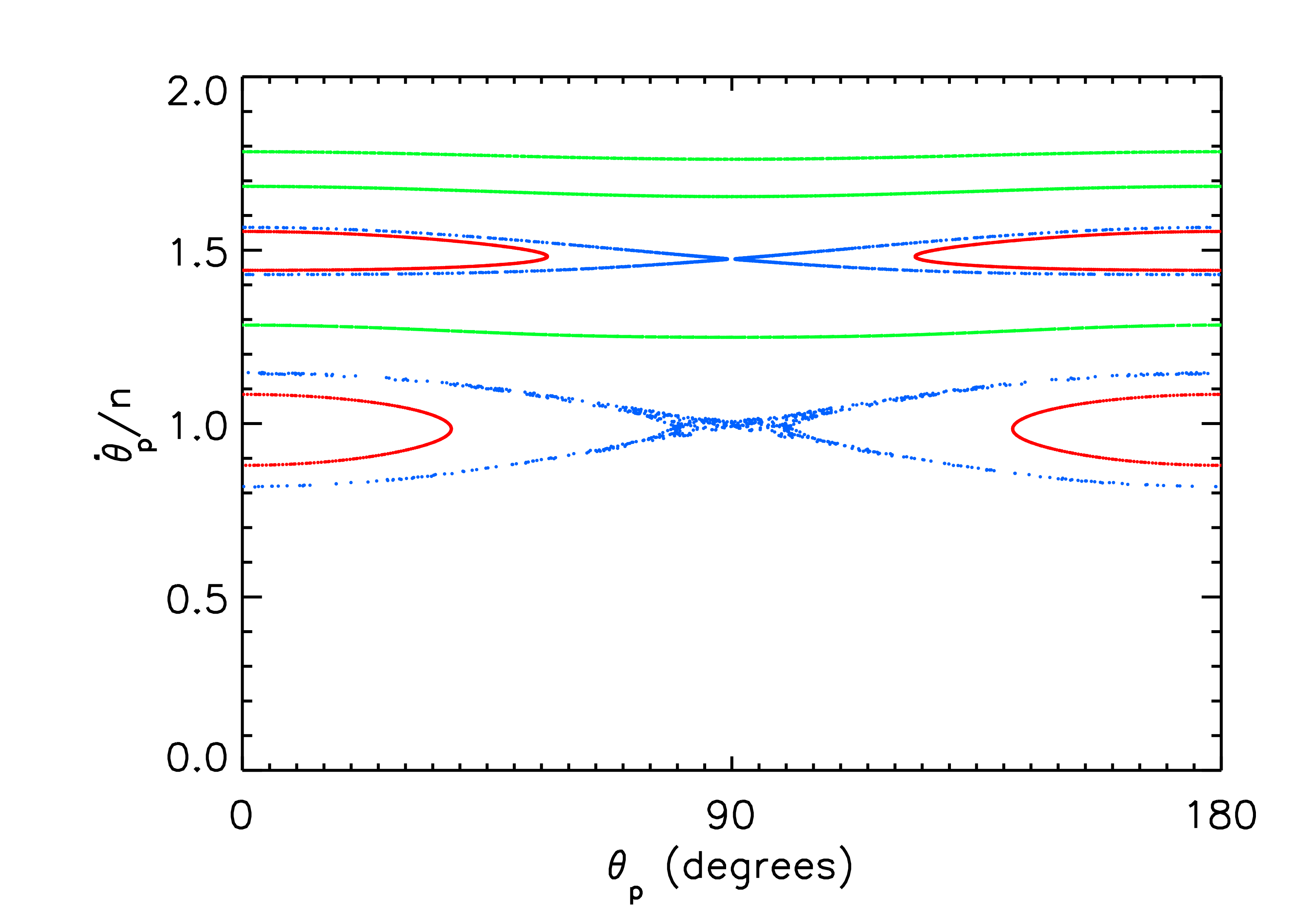}
\caption{
Surface of section plot for a secondary elongation $a/b=1.01$,
corresponding to $\omega_0=0.17$, and mutual orbit eccentricity
$e=0.05$.  Other system parameters are listed in
Table~\ref{tab:parameters}. 
Seven trajectories with initial
$\dot{\theta_p}/n$ values of 1.08, 1.15, 1.28, 1.55, 1.57, 1.68, and
1.78 are plotted. 
Initial $\theta_p$ values are 0 in all cases. Red,
blue, and green colors indicate resonant, chaotic, and non-resonant
quasi-periodic trajectories, respectively.\\}
\label{fig:ss_reg}
\end{figure}

It is easy to differentiate between regular and chaotic trajectories
on surface of section plots: regular trajectories fall on smooth
curves, whereas chaotic trajectories fill up an area of the phase
space over successive visits \citep{wisdom84}.
Figure~\ref{fig:ss_reg} shows different types of trajectories of a
slightly elongated secondary in this phase space. The system
parameters for this plot are based on radar-derived estimates for
near-Earth asteroid (NEA) 1991~VH~\citep{margot08, naidu12} and are
given in row 3 of Table~\ref{tab:parameters}.  The plot looks
symmetric about $\theta_p=90^\circ$ because we use triaxial ellipsoids
for the simulations, so $\theta_p=0^\circ$ is equivalent to
$\theta_p=180^\circ$. Seven trajectories with different initial
conditions are shown in this figure.  Throughout a simulation, the
secondary remains on the trajectory it started on. The red and green
trajectories are regular quasi-periodic, whereas the blue trajectories
are chaotic.

On a resonant (red color) trajectory, the secondary librates in a
spin-orbit resonance region.  For the red trajectory surrounding
$\dot{\theta_p}/n=1.5$, the secondary spins three times for every two
orbits, so it is in a 3:2 spin-orbit resonance. Mercury is the only
known object in a 3:2 spin-orbit resonance.  For the red trajectory
surrounding $\dot{\theta_p}/n=1$, the secondary is in a 1:1 spin-orbit
resonance, i.e., it spins synchronously (e.g., the Earth's
moon). 
Similar trajectories with islands centered exclusively on
$\theta_p=0^\circ$ and $\theta_p=180^\circ$ exist near half-integer
values of $\dot{\theta_p}/n$ (2, 2.5, 3, etc.).  The horizontal extent
of the trajectory around $\theta_p=0^\circ$ gives the amplitude of
excited-mode libration (equivalent to free libration in the decoupled
spin problem). For example, on the red trajectory in the 1:1 resonance
region, the secondary has an excited mode libration amplitude of
$\sim37^\circ$. A trajectory with only relaxed-mode libration plots as
a point on the y-axis, which we call
the {\em center} of the resonance region (not shown in the figure).  The
relaxed-mode libration is not detectable in the horizontal dimension
of the surface of section plots because we sample the spin state of
the secondary at pericenter, where the relaxed-mode libration is
always at $0^\circ$ phase.  However, the relaxed-mode libration is
detectable in the vertical dimension of the surface of section plots
because it contributes to the angular velocity of the secondary at
pericenter.  The centers of the resonance regions are displaced
vertically from their nominal positions in the absence of relaxed-mode
libration. These offsets can be seen clearly for relaxed-mode
librations with larger amplitudes (Figures \ref{fig:vh_ss} and
\ref{fig:yt1_ss}).  
They are strictly due to torques on the permanent deformation of the
satellite and are unrelated to the tidally induced pseudo-synchronous
rotation described by, e.g., \citet{ferrazmello13} for nearly
spherical satellites on eccentric orbits.

A chaotic (blue color) trajectory marks the boundary of a resonance
region and is called a separatrix. On a separatrix the secondary
explores the entire range of $\theta_p$ values and the trajectory
fills up a region of phase space, indicating that the trajectory is
chaotic.

Secondary spin rates that are further away from the resonance regions
put the secondary on a trajectory similar to one of the non-resonant
quasi-periodic (green) trajectories. On these trajectories the
secondary is not in a spin-orbit resonance and circulates through all
$\theta_p$ values in a quasi-periodic manner.  These trajectories are
called quasi-periodic because they exhibit at least one
non-commensurate frequency in addition to the frequency at which the
motion is sampled.

\citet{wisdom84} assumed that the secondary spin is decoupled from the
mutual orbit, a reasonable assumption for the Saturn-Hyperion system
because Hyperion has negligible angular momentum compared to the
mutual orbit. Under this approximation, they derive the half-widths of
the resonance regions (equation~(\ref{eq:res_width}), in units of the
mean motion) and of the chaotic separatrix surrounding the 1:1
spin-orbit resonance (equation~(\ref{eq:chaos}), in the energy domain):

\begin{equation}
\frac{1}{2}RW=\omega_0\sqrt{|H(p,e)|};
\label{eq:res_width}
\end{equation}

\begin{equation}
\frac{1}{2}SW = \frac{\Delta E}{E_0} \approx \frac{14\pi e}{\omega_0^3} e^{-(\pi/2\omega_0)}.
\label{eq:chaos}
\end{equation}
\noindent Here, $H$ are functions tabulated in~\citet{cayley1861} and
$p$ is the spin-orbit resonance ratio, e.g., $p=3/2$ for a 3:2
spin-orbit resonance. $\Delta E$ represents energy variations on the
chaotic separatrix and $E_0$ comes from the first integral of the
averaged equation of motion of libration \citep{wisdom84}.  It is the
energy at which the libration angle begins to circulate:
\begin{equation}
E_0=\frac{1}{4}n^2\omega_0^2C,
\label{eq:separatrix}
\end{equation}
where $C$ is the moment of inertia about the spin axis.
In the averaged equation of motion, the higher frequency
terms that give rise to chaos are ignored, so the separatrix is
regular.

The width of the resonance and of the chaotic regions grow larger with
$\omega_0$ and $e$. For large enough values, neighboring resonance
regions overlap, resulting in a large chaotic zone surrounding the
overlapping resonances.  The resonance overlap criterion for the 1:1
and 3:2 spin-orbit resonances is given by \citep{wisdom84}:
\begin{equation}
\omega_0^{RO}=\frac{1}{2+\sqrt{14e}}.
\label{eq:ro}
\end{equation}
Overlap occurs when $\omega_0 > \omega_0^{RO}$.

\begin{figure}
\plotone{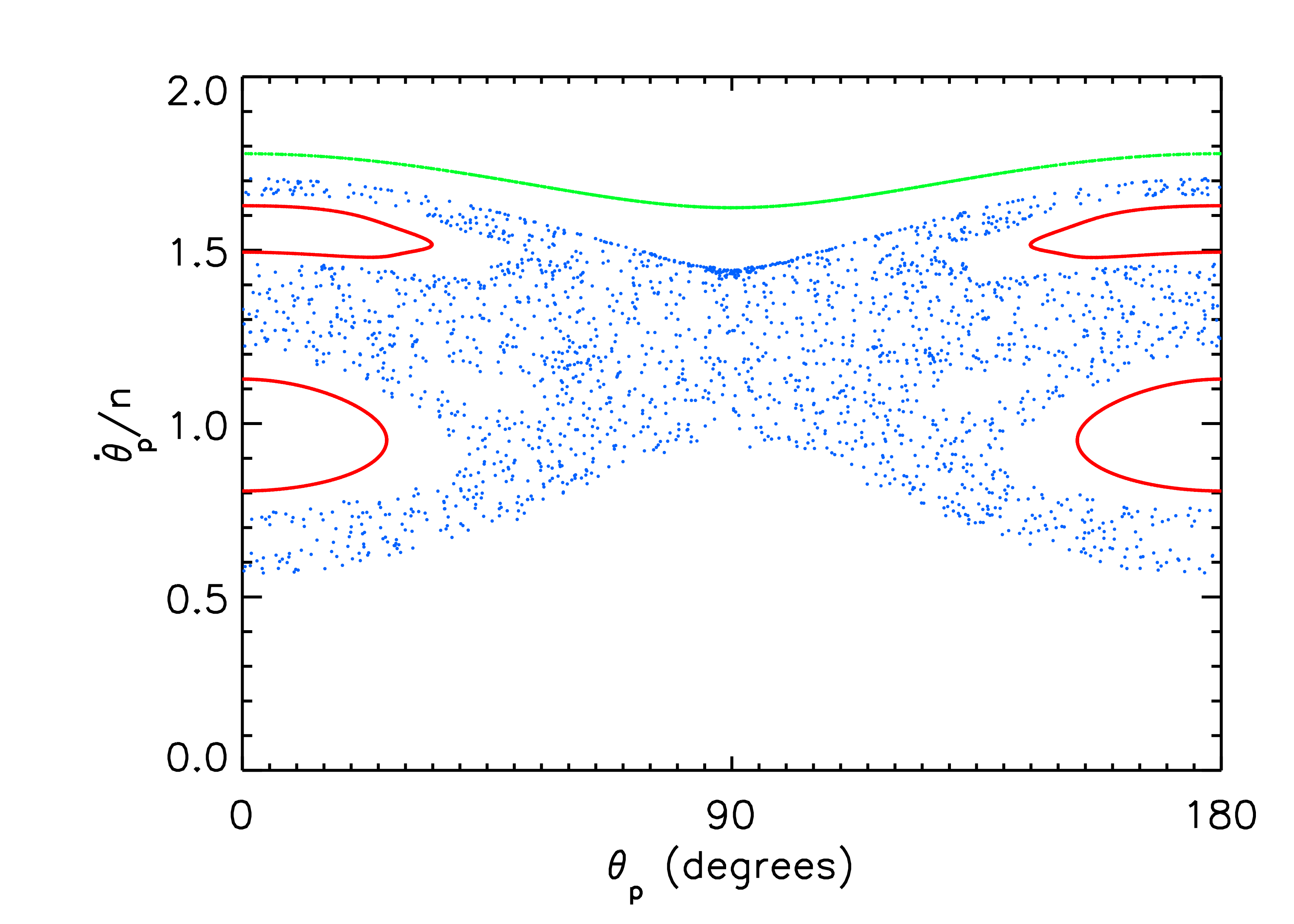}
\caption{Surface of section plot illustrating the partial overlap of
the 1:1 and 3:2 spin-orbit resonances for a secondary elongation
$a/b=1.06$, corresponding to $\omega_0=0.42$, and a mutual orbit
eccentricity $e=0.05$. Other system parameters are given in
Table~\ref{tab:parameters}. 
Four trajectories with initial
$\dot{\theta_p}/n$ values of 1.13, 1.63, 1.70, and 1.78 are
plotted.
Initial $\theta_p$ values are 0 in all cases. Color scheme as
in Figure~\ref{fig:ss_reg}.}
\label{fig:ss_overlap}
\end{figure}

These equations were derived under the assumption that the secondary
spin has no feedback on the mutual orbit, and we investigate whether
the analytical formulation (equation (\ref{eq:ro})) matches the
results of our coupled integrator.  For a system based on 1991~VH
(Table~\ref{tab:parameters}), we varied the elongation in steps of
0.01 and determined when resonance overlap occurred.  We find that it
does not occur for $a/b=$ 1.03 ($\omega_0=$ 0.30) but that it does
occur for $a/b \geq$ 1.04 ($\omega_0 \geq$ 0.34).  The analytical
estimate, which does not take the width of the separatrix into
account, places the onset of chaos at $\omega_0^{RO}=0.35$ for
$e=0.05$.  
The small difference between the analytical and numerical estimates
for the onset of chaos suggests that equation (\ref{eq:ro}) provides a
reasonable approximation even in the presence of spin-orbit coupling.
In subsequent sections, we will confirm this finding by providing
values for both estimates for a variety of orbital eccentricities.
Note that \citet{wisdom84} also observed a small difference between
analytical and numerical estimates, even in the fully decoupled case.
To illustrate resonance overlap, we generate a surface of section for
a value of $a/b=1.06$ and $e=0.05$ such that $\omega_0=0.42 >
\omega_0^{RO}=0.35$ (Figure~\ref{fig:ss_overlap}).
The overlap wipes out the non-resonant quasi-periodic trajectories
between the overlapping resonances and results in smaller 1:1 and 3:2
spin-orbit resonance regions and a large chaotic zone surrounding the
resonances.

\begin{figure}
\plotone{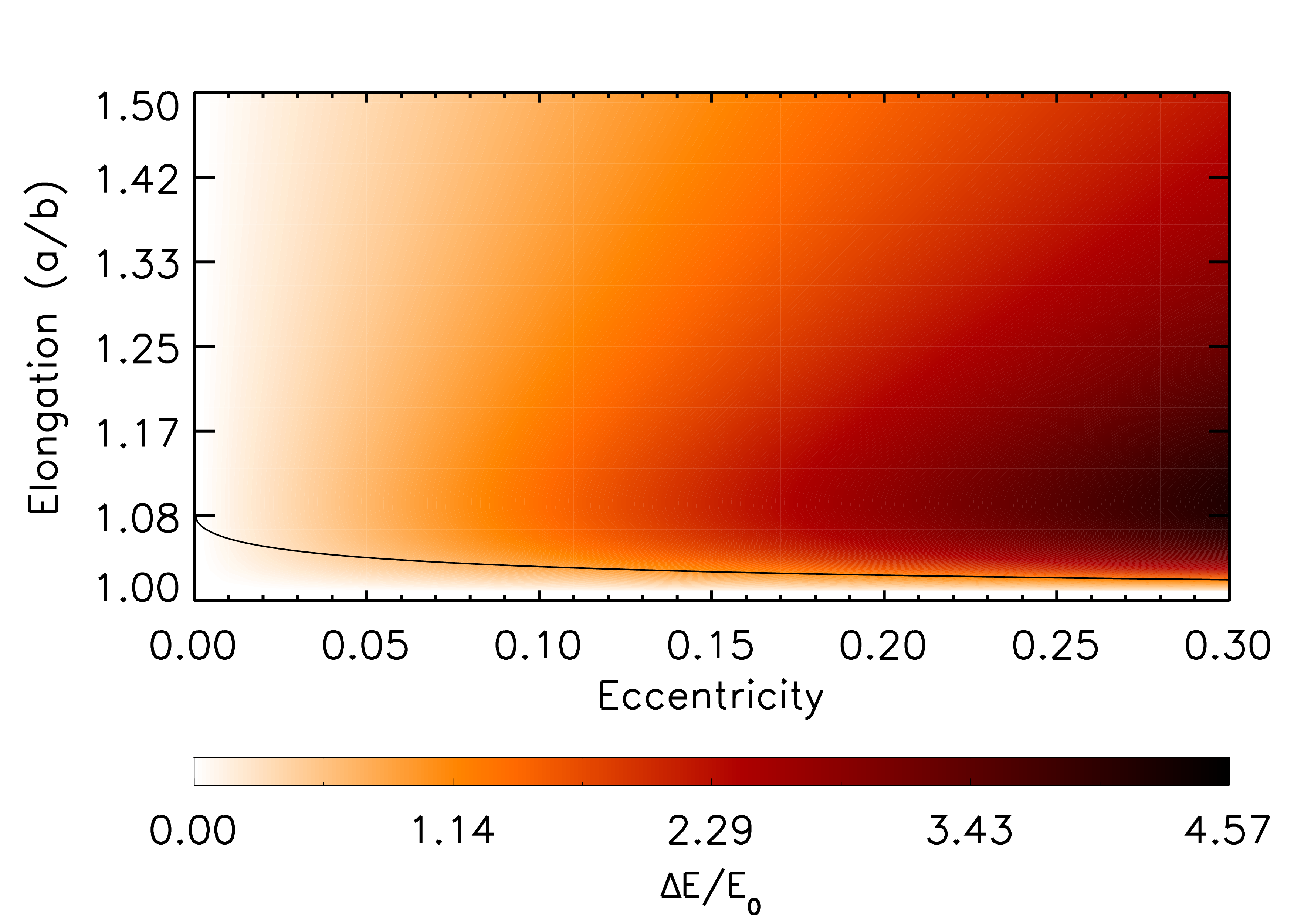}
\caption{Contour plot showing the half-width of the chaotic separatrix
  $\Delta E/E_0$ as a function of $a/b$ and $e$
  (equation~(\ref{eq:chaos})).
According to the resonance overlap criterion of
equation~(\ref{eq:ro}), overlap will occur in systems that lie above
the solid line. }
\label{fig:ro_chaos}
\end{figure}

Substituting $e=0$ in equation~(\ref{eq:ro}) yields
$\omega_0^{RO}=0.5$ which corresponds to $a/b\approx1.09$.  This value
is low compared to typical elongations observed in
asteroids~\citep[e.g.,][]{hudson95, hudson00, naidu13}. The secondary
of 1999~KW4 has an elongation of 1.3~\citep{ostro06}. This suggests
that resonance overlaps are quite likely to happen in binary
near-Earth asteroids.  However, for small eccentricities the width of
the chaotic separatrix remains small as dictated by
equation~(\ref{eq:chaos}), so resonance overlaps do not result in
large chaotic regions.  The resonance overlap threshold of $a/b$ as a
function of $e$ (equation~(\ref{eq:ro})) and the width of the
chaotic separatrix ($\Delta E/E_0$) as a function of $e$ and $a/b$
(equation~(\ref{eq:chaos})) are plotted in
Figure~\ref{fig:ro_chaos}.  The figure illustrates that the size of
the chaotic zone increases with eccentricity.

In the next section, we examine the surface of section plots for
well-characterized binary and triple systems.

\section{Well-Characterized binary and triple NEA Systems}
\label{sec:systems}

We simulate the spins and orbits of well-characterized binaries and
triples listed in \citet{fang12}, which includes both {\em
  synchronous} ($\langle \dot{\theta}/n \rangle=1$) and {\em
  asynchronous} ($\langle \dot{\theta}/n \rangle \neq 1$)
systems, 
where $\langle.\rangle$ indicates values averaged over one orbit.
We use our integrator to determine the minimum elongation at which
resonance overlap occurs and, for synchronous satellites, the
amplitude of relaxed-mode libration.  We also plot surfaces of section
for each system to examine the variety of dynamical regimes.  When
satellite elongations are not known, we assume a value of 1.3, which
corresponds to that of the 1999 KW4 satellite.  Equivalent radii for
the components and mutual orbital parameters are obtained from
\citet{fang12}, unless otherwise indicated.

In all cases, we assume the primaries to be spherical and the
secondaries to be triaxial ellipsoids.  We need a prescription for
choosing the axial dimensions such that they conform to the 
radius and mass of the secondary described in the literature.  $a$ and
$b$ are chosen to satisfy two conditions: 1) $a \times b=R_s^2$, where
$R_s$ is the radius of the secondary, and 2) $a/b$ equals the desired
elongation.  $c$ is chosen in a way that ensures $A<B<C$.  The choice
of $c$ is not crucial because the dynamics are mostly sensitive to the
value of $\omega_0=\sqrt{3(B-A)/C}$ which, for a triaxial ellipsoid,
is equal to $\sqrt{3(a^2-b^2)/(a^2+b^2)}$). 
The adopted
density of the satellite is
based on the observed mass of the satellite 
and on the volume of the triaxial ellipsoid.
Because the choice of $c$ is arbitrary, the densities used in the
simulations are not identical to the nominal densities, but the masses
used in the simulations do conform to the nominal masses.
We verified the robustness of our results by running simulations with
up to 20\% changes in the values of $c$ and found no appreciable
difference in the surface of section plots.

\subsection{1991 VH}
\label{sec:vh}
\begin{figure}
\plotone{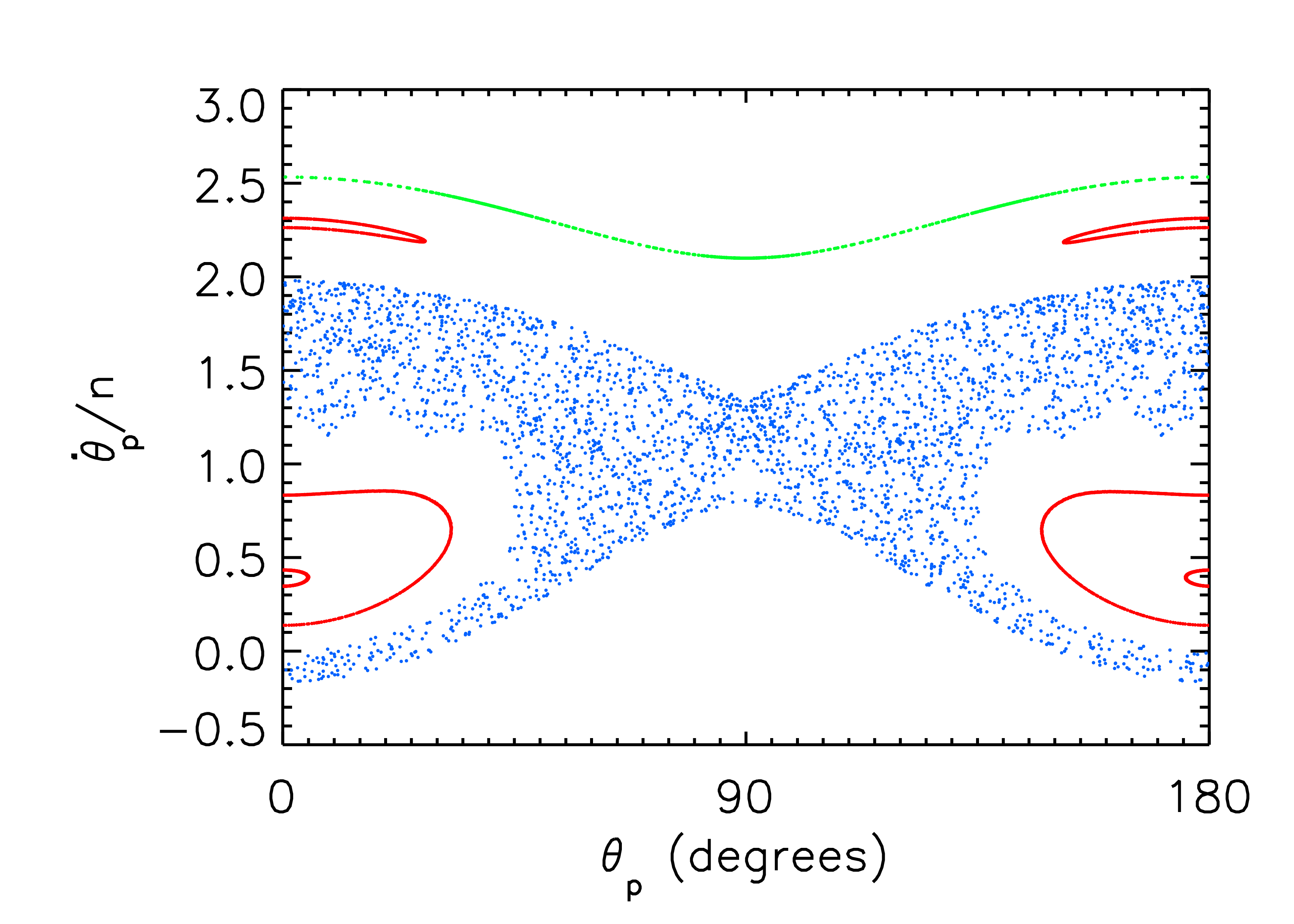}
\caption{Surface of section plot for the 1991~VH secondary using the
radar-derived secondary elongation $a/b=1.5$, corresponding to
$\omega_0=1.07$, and the mutual orbit eccentricity $e=0.05$. Other
system parameters are given in Table~\ref{tab:parameters}.  
Five
trajectories with initial $\dot{\theta_p}/n$ values of 0.43, 0.83, 1.63,
2.31, and 2.53 are plotted.
Initial $\theta_p$ values are 0 in all
cases. Color scheme is the same as in figure~\ref{fig:ss_reg}.\\}
\label{fig:vh_ss}
\end{figure}

As mentioned in section~\ref{sec:sos}, the overlap of the 1:1 and 3:2
spin-orbit resonances of
the 1991 VH secondary happens for ${a/b>1.04}$.  Radar images show
that its equatorial elongation is about
1.5~\citep{naidu12}. Figure~\ref{fig:vh_ss} shows a surface of section
plot for $e=0.05$ and $a/b=1.5$ (other parameters
are listed in Table~\ref{tab:parameters}). At these values, the
chaotic zone completely wipes out the 3:2 
spin-orbit resonance
but a large stable 1:1 spin-orbit resonance region still
exists.  The center of the sychronous region (as defined in
section~\ref{sec:sos}) is on the y-axis in Figure~\ref{fig:vh_ss}, in
the region bounded by the smaller red trajectory close to
$\dot{\theta_p}/n=0.5$. It is shifted down from $\dot{\theta_p}/n=1$
due to relaxed-mode libration which makes a non-zero contribution to
$\dot{\theta}$ at pericenter.  We measure the relaxed-mode libration
amplitude at the resonance center to be about 35$^\circ$.

The synchronous region is surrounded by a chaotic zone.  This has
implications for synchronous capture that are discussed in
section~\ref{sec:implications}.  If the secondary gets captured in the
synchronous region, tides are expected to damp the excited-mode
libration of the secondary, driving its trajectory towards the center
of the synchronous region, where it exhibits only relaxed-mode
libration.  
Since the spin is coupled to the orbit, energy removed from the
secondary spin will gradually change the orbit, the surface of section
map, and the relaxed-mode libration amplitude and frequency.
Throughout this evolution, the secondary remains in the same dynamical
regime close to the center of the synchronous region.
The next higher order stable resonance is the 
2:1 resonance, however probability of capture into this resonance is
low ($\sim10^{-3}$ using equation 5.110 of \citet{murray99}).
Similar to the synchronous region, the 2:1 resonance
region is shifted vertically from
$\dot{\theta_p}/n=2$.  The shift in this case is upwards because the
instantaneous satellite spin rate at pericenter is greater than its
orbit-averaged value of $2n$.

Preliminary measurements of the Doppler extents (or bandwidths) of the
secondary in radar images~\citep{margot08,naidu12} are consistent with
chaotic behavior, but because of the large amplitude libration at the
resonance center and corresponding spin rate variations
(Section~\ref{sec:implications2}), we cannot entirely rule out the
possibility of synchronous spin.

\subsection{2003 YT1}

\begin{figure}
\plotone{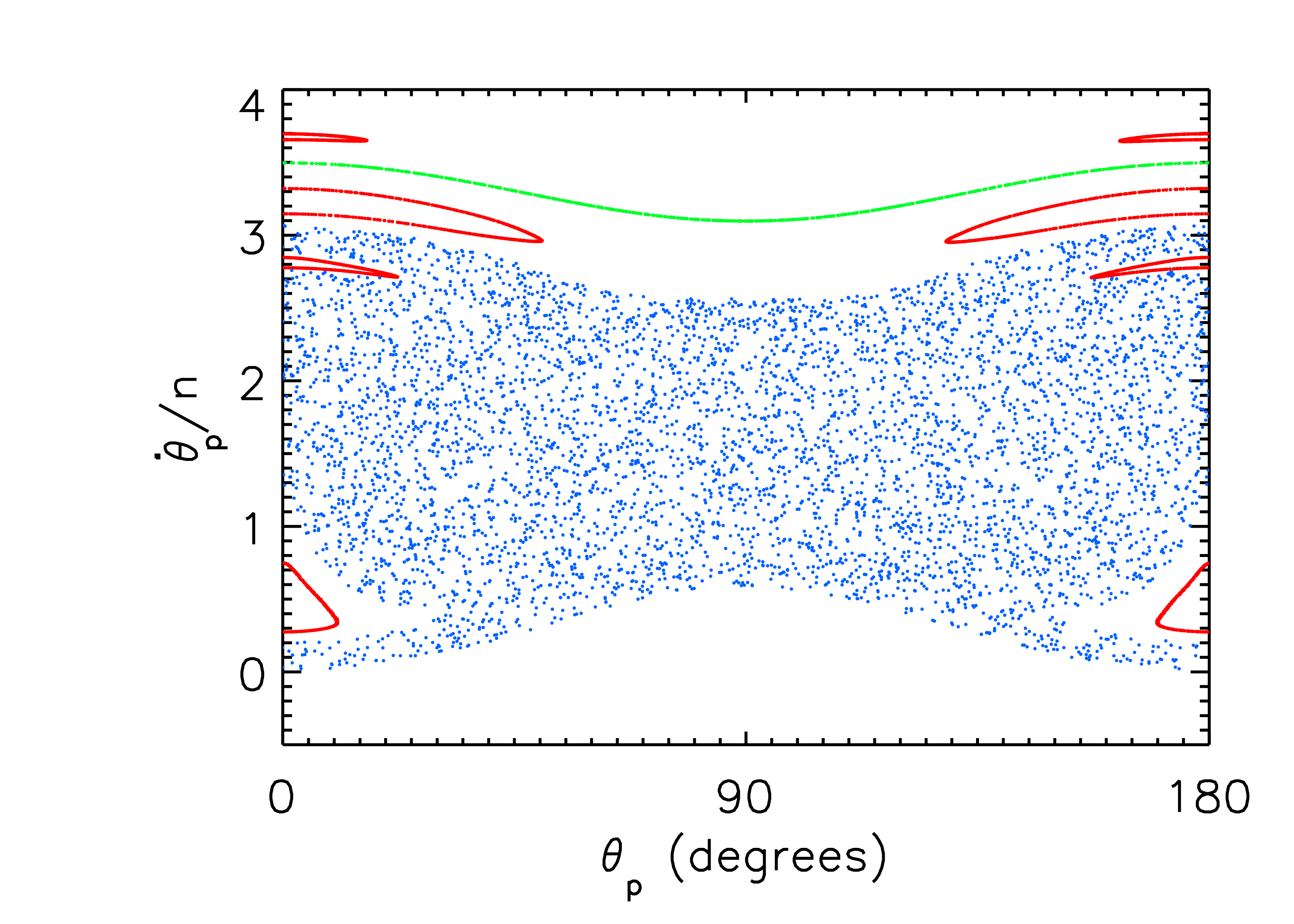}
\caption{Surface of section plot for the 2003~YT1 secondary assuming
the secondary elongation $a/b=1.3$, corresponding to $\omega_0=0.88$,
and the mutual orbit eccentricity $e=0.18$. Other system parameters
are given in Table~\ref{tab:parameters}. 
Six trajectories with
initial $\dot{\theta_p}/n$ values of 0.75, 1.29, 2.85, 3.15, 3.5, and 3.69 are
plotted.
Initial $\theta_p$ values are 0 in all cases. Color scheme is
the same as in Figure~\ref{fig:ss_reg}.}
\label{fig:yt1_ss}
\end{figure}

This system's component sizes are $R_p\approx550$~m and
$R_s\approx105$~m, so the primary is similar to that of 1991 VH but
the secondary is a few times smaller.  The orbit ($a/R_p \sim 7$) is
somewhat wider than that of 
1991~VH ($a/R_p \sim 5.5$), and it is also more eccentric ($e=0.18$
vs. $e=0.05$).  The smaller secondary and wider mutual orbit mean that
spin and orbit are less coupled in this system than in 1991~VH.
Substituting $e=0.18$ in equation~(\ref{eq:ro}), we get a theoretical
threshold for resonance overlap $\omega_0^{RO}=0.30$, which
corresponds to $a/b=1.03$.  Using our simulations we find that the
resonance overlap threshold lies between $\omega_0=0.24$ ($a/b=1.02$)
and $\omega_0=0.30$ ($a/b=1.03$).
The elongation of the satellite is unknown.  For our simulations
(Figure~\ref{fig:yt1_ss}), we chose an elongation of 1.3.

The 1:1 spin-orbit resonance region is not as prominent in this plot
as it is in Figure~\ref{fig:vh_ss} due to the higher
eccentricity. Despite the higher eccentricity, the y-axis location of
the synchronous region center is similar to that of 1991~VH, because
of the smaller satellite elongation. The chaotic region is much bigger
than that of 1991~VH and extends to $\dot{\theta_p}/n\approx3$.
The first higher order stable resonance is the 5:2 spin-orbit
resonance.  
This resonance region is shifted upwards from $\dot{\theta_p}/n=2.5$,
similar to 
the upward shift of the 2:1 resonance region of 1991~VH.
The amplitude of
the relaxed-mode libration measured at the center of the synchronous
region is about 45$^\circ$.

\subsection{2004 DC}

\begin{figure}
\plotone{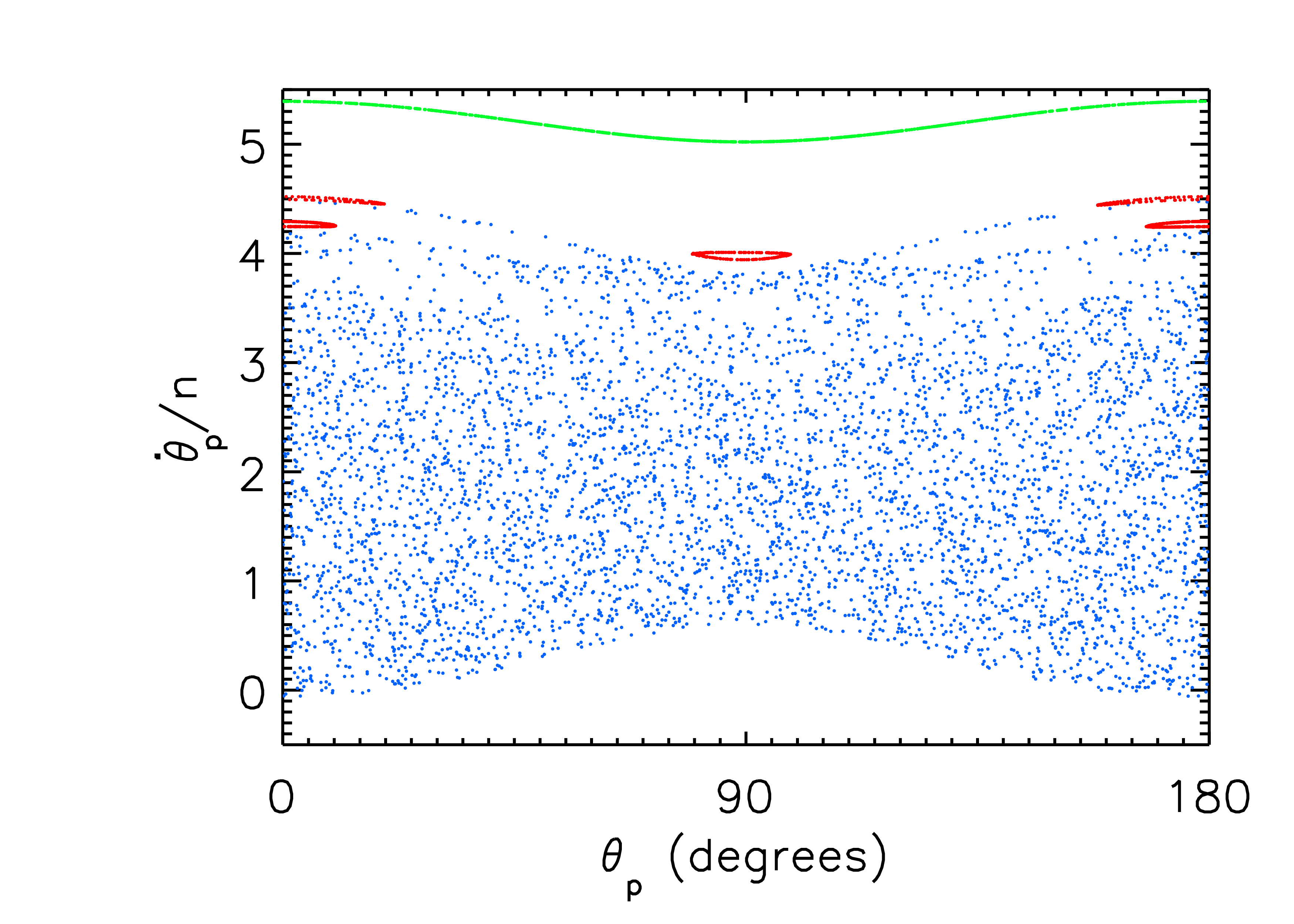}
\caption{Surface of section plot for the 2004~DC secondary assuming the
secondary elongation $a/b=1.3$, corresponding to
$\omega_0=0.88$, and the mutual orbit eccentricity $e=0.3$. Other system parameters are given in
Table~\ref{tab:parameters}. 
Four trajectories with initial $\dot{\theta_p}/n$ values of 1.39,
4.29, 4.49, and 5.39 are plotted.  
Initial $\theta_p$ values are 0 in
all cases. Color scheme is the same as in Figure~\ref{fig:ss_reg}.}
\label{fig:dc_ss} 
\end{figure}

2004~DC has the smallest primary ($R_p\approx180$ m), secondary
($R_s\approx30$ m), and mutual orbit semimajor axis (750 m) in our
sample, but it has the most eccentric mutual orbit ($e\approx0.3$)
(Table~\ref{tab:parameters}). The resonance overlap criterion
(equation~\ref{eq:ro}) gives $\omega_0^{RO}=0.25$, which corresponds
to an elongation of $a/b=1.02$. 
Using our simulations we find that the resonance overlap threshold
lies between $a/b=1.01$ ($\omega_0=0.17$) and $1.02$
($\omega_0=0.24$), roughly consistent with the analytical
estimate. The shape of the secondary is not known, however its
appearance in the radar images suggest that $a/b\leq1.3$ (Patrick
Taylor, personal communication).  Figure~\ref{fig:dc_ss} shows the
surface of section plot for a secondary having an elongation of 1.3
(system parameters in Table~\ref{tab:parameters}).  The chaotic region
is so large that even the 1:1 resonance region disappears and the
lowest-order stable resonance region is the 4:1 spin-orbit
resonance. In fact the synchronous island is absent for all values of
satellite elongations $\ge1.1$.

\subsection{Synchronous Satellites}

\begin{figure}
\plotone{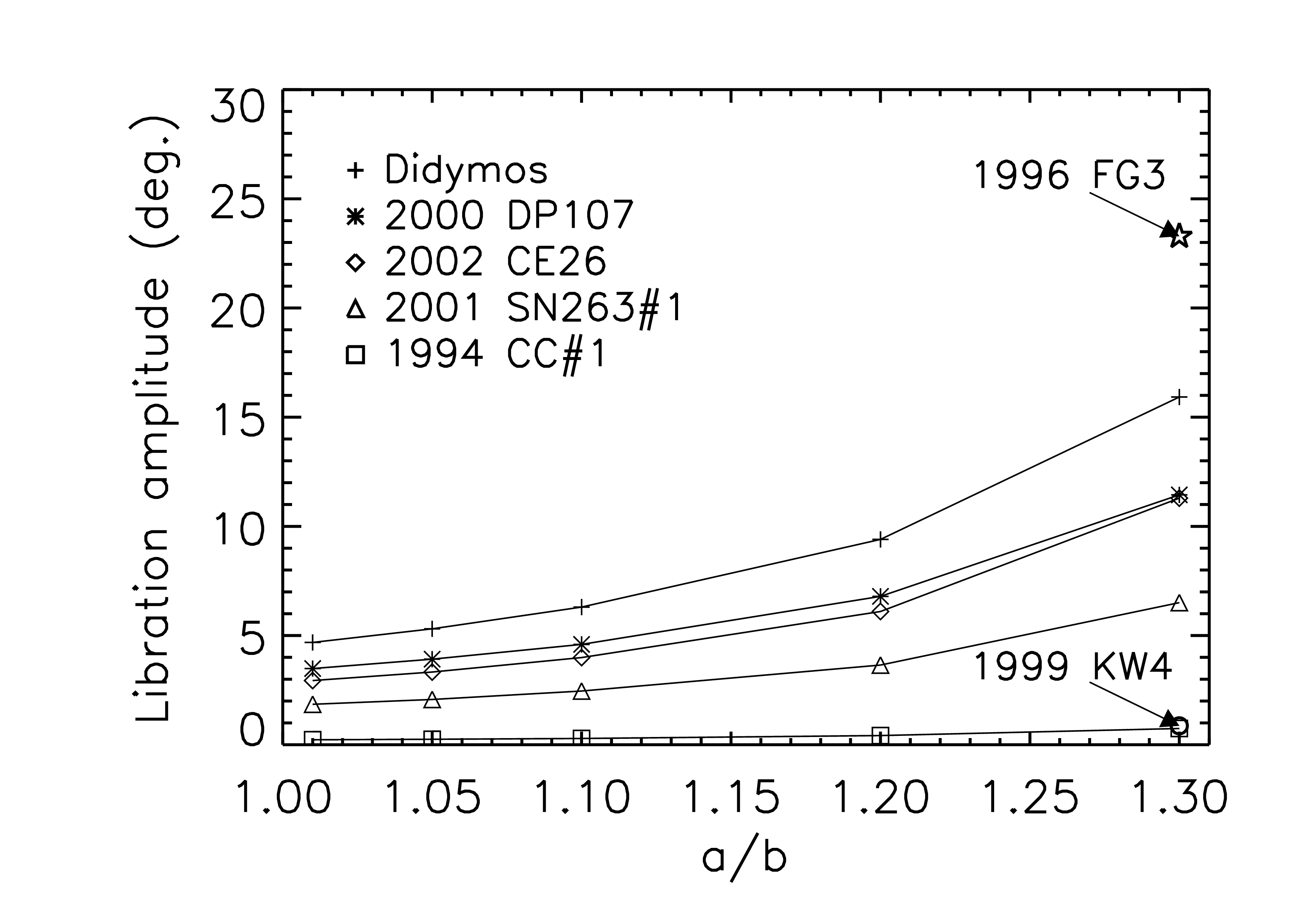}
\caption{Numerical estimates of relaxed-mode libration amplitude as a
  function of satellite elongation for synchronous satellites among
  well-characterized binary and triple systems.  For 1999 KW4 and 1996
  FG3, we plot single points corresponding to the known elongations of
  the satellites. The libration amplitude of 1996~FG3 is an upper
  limit based on an eccentricity of 0.07.}
\label{fig:all_lib}
\end{figure}

Radar data show that the satellites of 2000~DP107, 2002~CE26,
2001~SN263 (Gamma), 1999~KW4, and 1994~CC (Beta) are synchronous
\citep[respectively]{margot02,shepard06, nolan08,ostro06,brozovic11}.
Didymos may also be synchronous~\citep{benner10}; for our purposes we
assume that it is. \citet{scheirich15} found that 1996~FG3 is
synchronous. We use radar-derived mutual orbital parameters, component
radii, and component masses for simulating these systems. These
parameters are given in Table~\ref{tab:parameters}.
For satellites whose elongations are not well known, we 
perform simulations using $a/b$=1.01, 1.05, 1.1, 1.2,
and 1.3. Each simulation is performed with initial conditions that put
the trajectory at the center of the synchronous island.
We identify the center of the synchronous island by varying the values
of initial $\dot{\theta}$ until the horizontal extent of the
trajectory on the surface of section becomes $\sim$0.
As mentioned in
section~\ref{sec:vh}, the excited-mode libration amplitude is zero at
the center of the synchronous island, which is what is expected for a
tidally evolved satellite.  In this case, the satellite exhibits only
the relaxed-mode libration, which we measure as the angle between the
long axis of the secondary and the line joining the primary and
secondary centers of masses (In the decoupled terminology, this is the
optical+forced libration).  The libration amplitudes, i.e., 
the maximum values of the libration angles, are plotted
as a function of elongation in Figure~\ref{fig:all_lib}.
The analytical estimates of the libration amplitudes, assuming the
decoupled spin-orbit problem, are given by equation~(\ref{eq:lib}).

Since shape and spin state modeling are tied to each
other~\citep[e.g.,][]{ostro06,naidu13}, calculations such as those
shown in Figure~\ref{fig:all_lib} are useful for shape modeling of
asteroid satellites. These estimates are also useful for modeling
binary YORP torques~\citep{cuk05} on synchronous satellites. If a
system exhibits excited-mode libration in addition to the relaxed-mode
libration, the two librations will add up and create a beating
pattern.  Because the amplitudes and frequencies of excited-mode
libration can span a wide range of values, 3D reconstruction and
binary YORP modeling of dynamically excited satellites is complicated.

\section{Implications}
\label{sec:implications}

\subsection{Presence of chaotic regions and synchronous capture}
\label{sec:implications1}
In section~\ref{sec:systems}, we showed that resonance overlap is
likely to occur for asynchronous satellites in our sample and that
large chaotic zones are expected in their phase spaces.  This behavior
can be expected in other, similar systems.  Let us consider the
evolution of a satellite formed with a high initial spin rate such
that its trajectory in phase space is in or above the chaotic zone,
the situation expected for most satellites in the formation model of
\citet{jacobson11}.  It is possible for YORP to increase the spin rate
of the satellite, but we focus on the spin-down evolution under the
influence of tidal and YORP forces.  
Satellites that start above the chaotic region will most likely
encounter the chaotic zone on their way to the 1:1 spin-orbit
resonance region.
In the chaotic zone, the satellite is acted upon by tides, YORP, and
torques on its permanent shape, which cause the chaotic spin.  Since
the spin of the satellite is coupled to the mutual orbit, angular
momentum removed or added to the secondary spin by tides and YORP will
also affect the mutual orbit and cause the surface of section map of
the secondary to vary. However, for binary NEAs, the angular momentum
of the mutual orbit is much greater than that of the secondary spin,
so the effect is expected to be small. We neglect this effect for the
following discussion and assume the surface of section map to be
roughly constant during the evolution of the secondary spin in the
chaotic zone.

If the chaotic separatrix around the 1:1 spin-orbit resonance region
is extremely thin, as is the case for satellites having almost
spherical shapes, then the torques on the permanent shape will be
small, allowing tides or YORP to easily drive the satellite spin
across the separatrix and into the synchronous region.  For satellites
having larger chaotic zones, like the asynchronous satellites in
section~\ref{sec:systems}, tides or YORP cannot simply sweep the
satellite across the chaotic region because torques on the permanent
shape can increase as well as decrease the spin rate of the satellite
in a random manner. The synchronous capture process is essentially
stochastic in nature. For capture to occur, the satellite has to spend
enough time near the boundary of the synchronous region for tides or
YORP to torque the satellite into resonance. Such a process was
discussed by \citet{wisdom84}. The details of this capture process are
not known and are difficult to model, however the probability of this
happening will depend on the relative magnitude of $\Delta E/E_0$
compared to
the energy dissipated due to tides ($\delta E_{\rm tides}/E_0$) or YORP
($\delta E_{\rm YORP}/E_0$).

Simulations and equation~\ref{eq:chaos} show that values of chaotic
spin energy variations ($\Delta E/E_0$) for the asynchronous
satellites are within an order of magnitude of 1.  We estimate the
magnitude of tidal dissipation in one orbit using the following
equation from
\citet{murray99}:
\begin{equation}
\delta E_{\rm tides}=\pi\frac{3}{2}\frac{k_2}{Q}\frac{n^4}{G}R_s^5.
\label{eq:tides}
\end{equation}
Here $k_2$ is the love number, $Q$ is the tidal dissipation factor,
$n$ is the mean motion, $G$ is the gravitational constant, and $R_s$
is the radius of the secondary.  We approximate energy dissipation due
to YORP in one orbit by multiplying the YORP torque given in
\citet{steinberg11} by the satellite rotation over one orbit, $4\pi$
assuming 2 satellite rotations per orbit:
\begin{equation}
\delta E_{\rm YORP}=\frac{2 \pi R_s^3 L_{\odot} f_Y}{3cd^2\sqrt{1-e_{\odot}^2}}.
\label{eq:yorp}
\end{equation}
Here $L_{\odot}$ is the solar luminosity, $f_Y$ is the YORP torque
efficiency, $c$ is the speed of light, $d$ and $e_{\odot}$ are the
semimajor axis and eccentricity of the heliocentric orbit,
respectively.  In order to compare the tidal and YORP energy
dissipation with $\Delta E/E_0$, we normalize $\delta E_{\rm tides}$ and
$\delta E_{\rm YORP}$ using $E_0$ from equation~\ref{eq:separatrix}.

For computing $\delta E_{\rm tides}$ we assume $Q=100$ and estimate
$k_2$ values using three different models. In the rubble pile model of
\citet{goldreich09}, $k_2=10^{-5}R_s$, where $R_s$ is in km.  Using
the system parameters from Table~\ref{tab:parameters}, we determine
$\delta E_{\rm tides}/E_0$ for all the asynchronous satellites to be
between $10^{-9}$ and $10^{-8}$. Assuming the monolith model of
\citet{goldreich09} for the secondary yields lower values of $\delta
E_{\rm tides}$ because a monolith is more rigid than a rubble pile of
the same size and has a lower value of $k_2$.
\citet{jacobson11a} derived a different relation between love number
and radius, $k_2=2.5\times10^{-5}R_s^{-1}$, by assuming that orbits of
observed synchronous asteroid satellites are in an equilibrium state
such that tidal torques balance binary YORP torques.  Subsituting
$k_2$ values from this relation in equation~(\ref{eq:tides}) yields
$\delta E_{\rm tides}/E_0$ between $10^{-7}$ and $10^{-5}$.
We compute $\delta
E_{\rm YORP}$ by assuming $f_y=5\times10^{-4}$, the estimated value
for asteroid YORP \citep{taylor07,lowry07}. $\delta E_{\rm YORP}/E_0$
values for 1991~VH, 2003~YT1, and 2004~DC are $5\times10^{-7}$,
$3\times10^{-6}$, and $2\times10^{-5}$, respectively.

Unknown values of $Q$, $k_2$, and $f_y$ introduce uncertainties of a
few orders of magnitude in $\delta E_{\rm tides}$ and $\delta E_{\rm YORP}$
but these energy dissipation values are several orders of magnitudes
smaller than $\Delta E/E_0$, suggesting that chaotic variations in
energy dominate tidal and YORP dissipations in these systems. This may
substantially delay spin synchronization and, therefore, BYORP-type
evolution.

If the timescale for synchronous capture is long, then tides may damp
the mutual orbit eccentricity significantly before spin
synchronization. This will reduce the size of the chaotic zone as
dictated by equation~\ref{eq:chaos} and make it easier for tides or
YORP to torque the secondary into the synchronous region.  Tidal
damping of eccentricity is not a very effective process and timescales
may be quite long.  \citet{fang12} estimated timescales in the range
$10^7$ to $10^{10}$ years for the asynchronous satellites, but these
may be in error because the underlying formalism by
\citet{goldreich63} assumes synchronous rotators.  It is likely that
energy dissipates faster in the case of satellites that are torqued
and tidally deformed in a chaotic manner, but the nature and
characteristic timescale of the eccentricity evolution remain poorly
known.  Complicating the picture is the fact that other mechanisms
such as solar perturbations~\citep{scheeres06} or planetary
flybys~\citep{fari92b,fang12encounters} may also be effective at
damping or exciting eccentricities.  Although there is uncertainty
related to the eccentricity-damping timescale, BYORP-type evolution
cannot take place until the spin period is synchronized to the orbital
period.  Asteroid binaries may enjoy extended lives because their
chaotically spinning secondaries prevent BYORP evolution.

\subsection{Interpretation of observational data}

\label{sec:implications2}

Our results have implications for radar and lightcurve data
interpretation. In radar observations (images and spectra), the
Doppler extent (or bandwidth) of an object is proportional to its
apparent, instantaneous spin rate (inversely proportional to its spin
period).  Lightcurves show variations in the object's brightness as it
spins. If the object is spinning at a constant rate, the brightness
variations will be approximately periodic.  The primary periodicity in
the lightcurve \citep[e.g.,][]{pravec06} is often used as a proxy for
the object's spin rate, even though the signal is affected by changes
in relative positions between the Sun, the object, and the observer.
We showed in section~\ref{sec:systems} that asteroid satellite spin
rates can be time-variable.  When the lightcurve data are of
sufficient quality and when $R_s/R_p\gtrsim0.2$, it is sometimes
possible to distinguish the signal of the secondary from that of the
primary.  In radar data, where the secondary is typically easily
detectable, the spinning satellite will exhibit approximately periodic
bandwidth variations.  However, in both cases, a variable spin rate
severely complicates the analysis.  Understanding the time-varying
nature of the satellite spin is important when analyzing radar and
lightcurve data.

\begin{figure}
\plotone{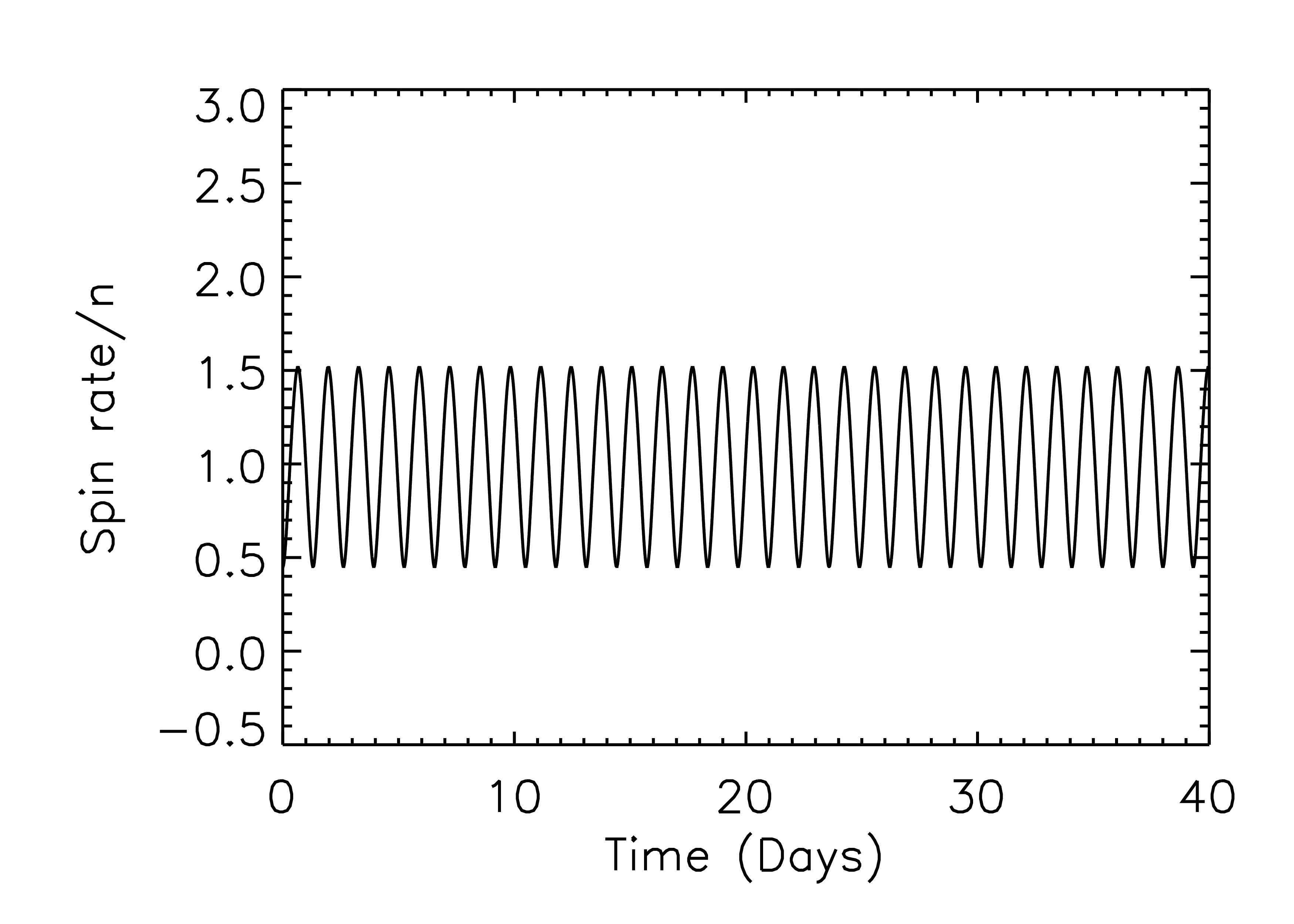}
\plotone{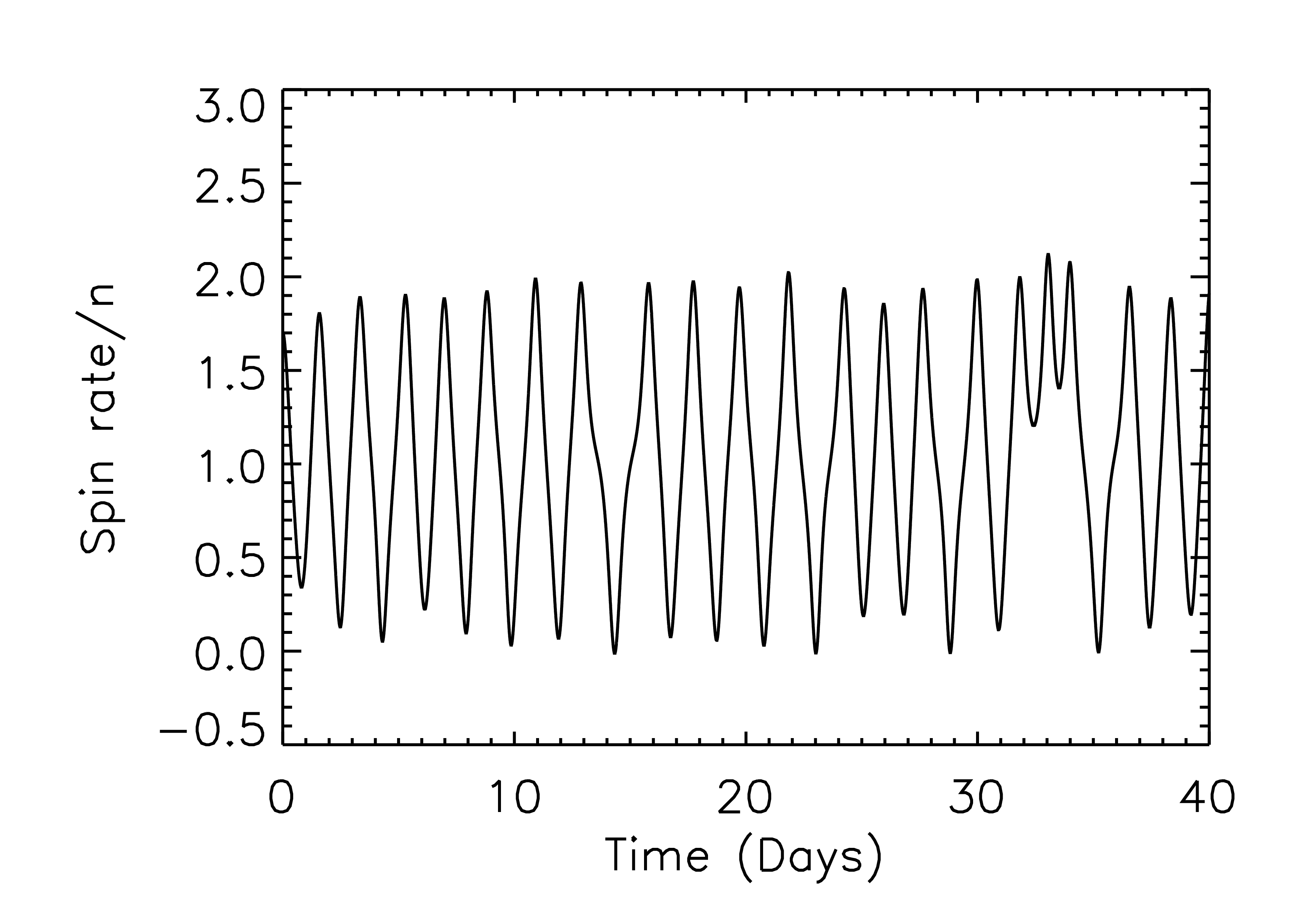}
\plotone{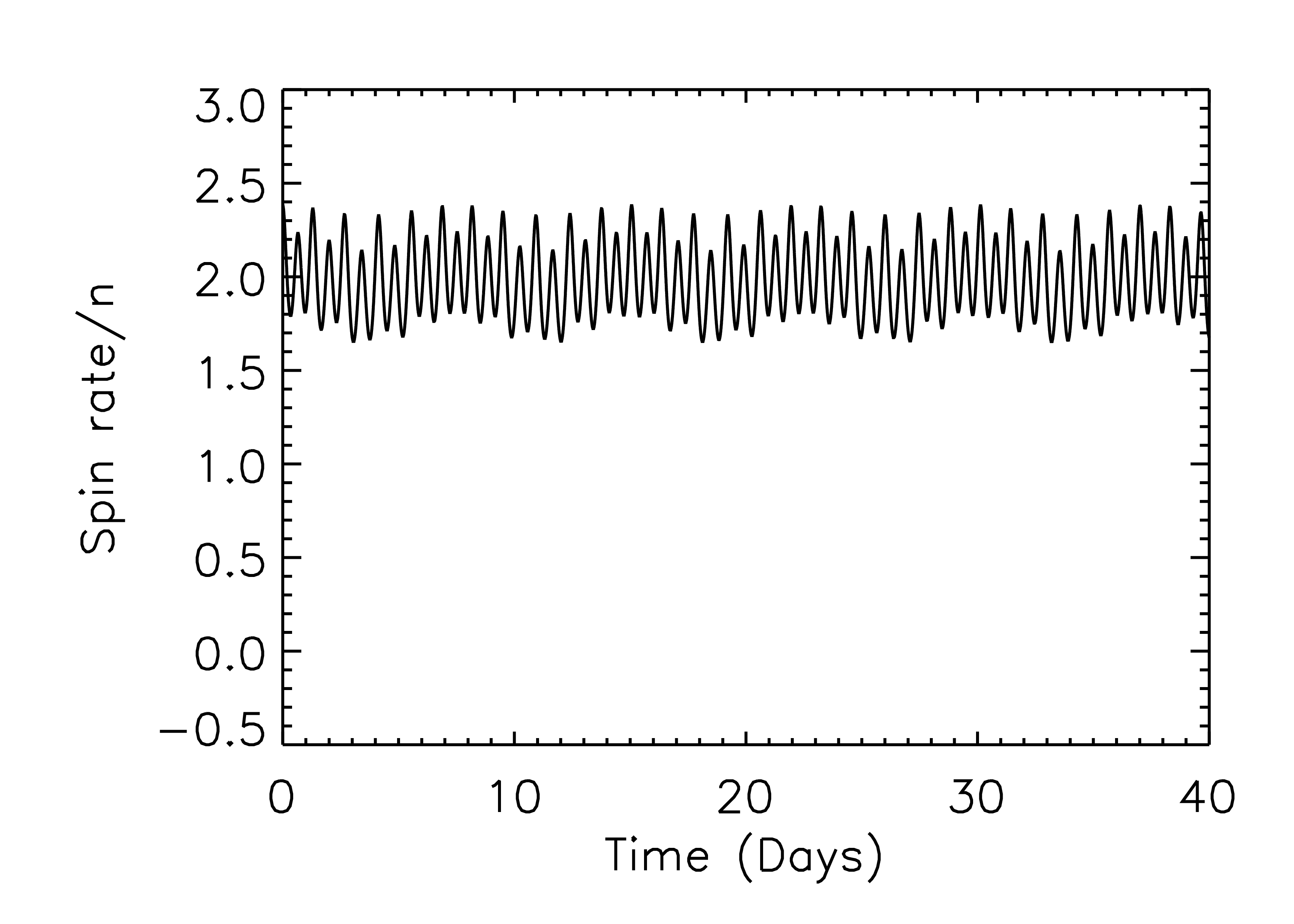}
\plotone{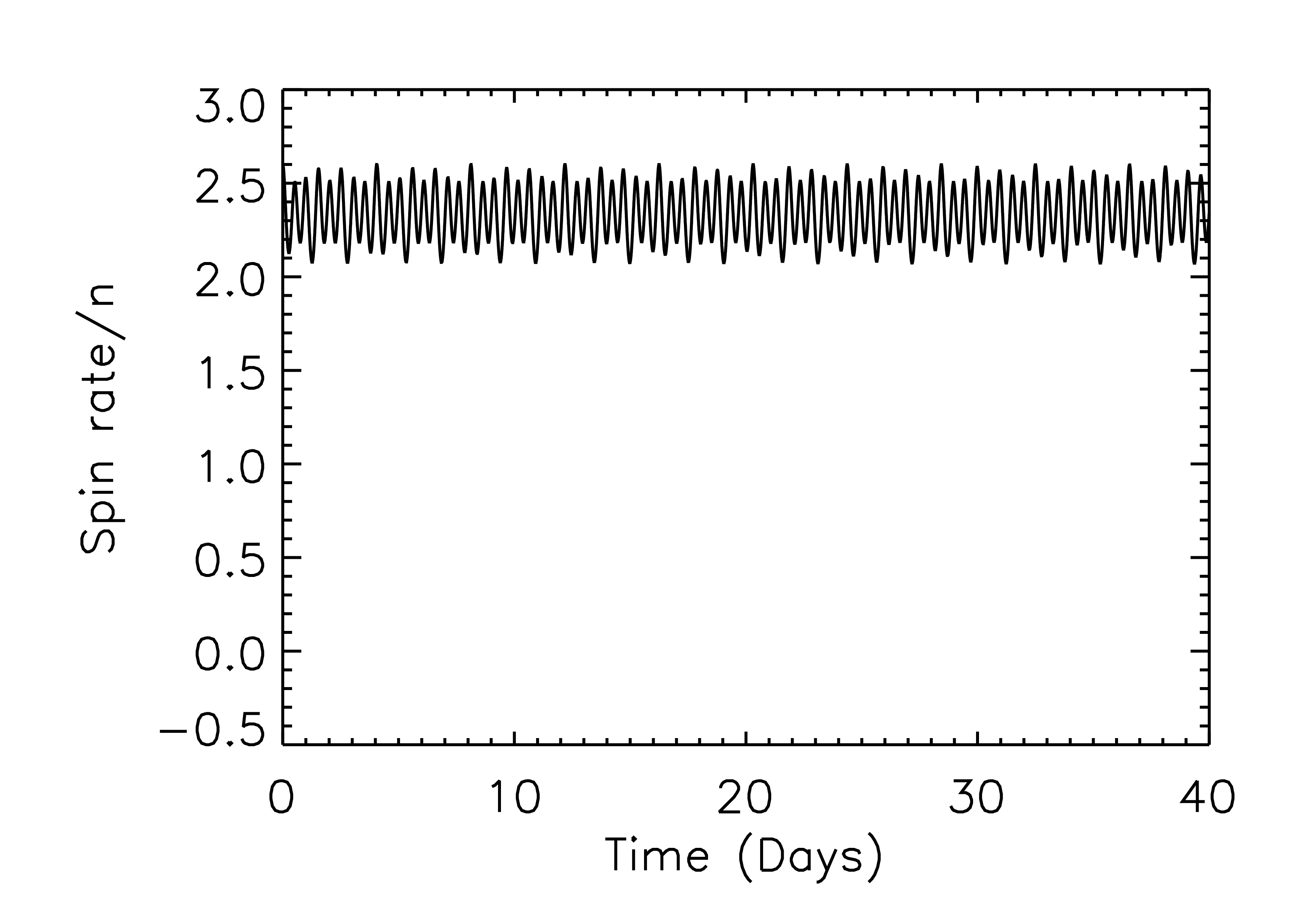}
\caption{Satellite spin rate variations for 4 possible trajectories of
1991~VH. 
From top to bottom, the initial 
values
of satellite spin rate, normalized by $n$,
are
0.45, 1.70, 2.38,2.60.
}
\label{fig:vh}
\end{figure}

Figure~\ref{fig:vh} shows spin rate as a function of time for four
trajectories of 1991~VH. The top panel shows a trajectory at the
center of the synchronous island. Even though it would plot as a point
on a surface of section, the satellite spin rate exhibits a large
oscillation at the orbital rate with an amplitude slightly greater
than the mean motion, $n$. As mentioned in section~\ref{sec:sos}, this
oscillation is the relaxed-mode libration of the satellite.

The second panel in Figure~\ref{fig:vh} shows a chaotic trajectory.
The spin rate variations span a similar, but slightly larger, range of
values than that in the synchronous case.  If observations (radar or
photometric) were sparse, it would be difficult or even impossible to
ascertain whether a trajectory was periodic or chaotic.  With a
sufficient number of data points sampled at a sufficiently fast
cadence, one could examine the distribution of spin rate values to
identify the type of trajectory, as the distributions for resonant and
chaotic trajectories are different.  Spin rate variations on the
synchronous trajectory resemble a sinusoid, so the distribution of
spin rates looks approximately bimodal.  The spin rate distribution of
the satellite in the chaotic region cannot be generalized and depends
on specific system parameters.

The third and fourth panels show the 3:2 resonant trajectory (red) and
the quasi-periodic (green) trajectory of Figure~\ref{fig:vh_ss},
respectively. In these cases, the variations are much smaller than the
previous two trajectories.  These rotational regimes are easier to
identify because large-scale chaotic variations are not present.

In radar data analysis, modeling the spin state and shape of objects
are tied to each other \citep[e.g.,][]{ostro06,naidu13}. Incorrect spin
state assumptions may yield incorrect shape models.  Our results
indicate that in some cases it will be impossible to identify the spin
state of the satellite, whereas in other cases an appropriate
rotational model will provide a good fit to the data.  A simple model
of uniform rotation may not be sufficient, as libration amplitudes can
cause displacements that exceed the image resolution.  We recommend
using a model that includes librations for the 3D reconstruction of
asteroid satellites with even moderate eccentricity and elongation.

\section{Conclusions}

We examined the rotational regimes of asteroid satellites using
surfaces of section.  The trajectories can be broadly classified as
resonant, non-resonant quasi-periodic, and chaotic.
In order to identify the specific type of spin
behavior, a dense time sampling of the satellite spin state is
necessary (section~\ref{sec:implications2}), however such datasets are
seldom available. 
Even densely sampled lightcurves, for instance, do not yield
measurements of the instantaneous spin state due to the necessity of
observing $\sim$1 full period to estimate the spin period.  In
section~\ref{sec:implications2}, we showed that even synchronous
satellites
can undergo large variations in spin rates, 
potentially masquerading as asynchronous satellites.  
Careful analysis of the data along with coupled spin-orbit simulations
can be used to correctly identify the spin behavior.
Identifying the spin configurations is essential 
for 
obtaining accurate physical models of the satellites.

The spin configurations of satellites play a crucial role in the
secular evolution of binary/triple systems under the influence of
forces such as tides and binary YORP.  For example, the binary YORP
torque acts only on satellites whose spin periods are integer
multiples of their orbital periods~\citep{cuk05} and some estimates
suggest that this torque could disrupt binary systems in just a few
tens of thousands of years~\citep{cuk10, mcmahon10}. Thus
understanding the process of spin synchronization is essential for
understanding the evolution of binaries. In
section~\ref{sec:implications1}, we showed that satellites may have
significantly longer spin synchronization timescales than those
estimated by considering tidal and/or YORP forces only.  This would
increase the fraction of asynchronous binaries in the observed
population beyond what one would expect on the basis of tidal
despinning timescales.  The corresponding delay in the onset of binary
YORP implies that the lifetimes of binary asteroids can be
significantly longer than the few tens of thousands of years suggested
by binary YORP models.

\section{Future Work}
We examined the results of spin-orbit coupling in the planar case.
However, \citet{wisdom84}, using 3D simulations, showed that seemingly
stable configurations in planar simulations 
can be attitude unstable. Future work will involve studying
inclined/oblique binary systems in order to test the attitude
stability of satellites in various regions of phase space.  Our
integrator can also be used for studying the secular evolution of
binary asteroids. This will require implemention of radiation pressure
and tidal forces.

\section*{Acknowledgments}
We thank Jack Wisdom, Dan Scheeres, Jay McMahon, and Seth Jacobson for
useful discussions, and the anonymous reviewer for excellent
suggestions.  This material is based upon work supported by the
National Science Foundation under Grant No. AST-1211581.

\bibliographystyle{plainnat}


\end{document}